\documentclass{aa}

\usepackage{graphicx}
\usepackage[colorlinks=true, linkcolor=blue, citecolor=blue]{hyperref}

\usepackage{txfonts}

\providecommand\linenumbers{}     
\providecommand\nolinenumbers{}   

\makeatletter
\@ifundefined{nolinenumbers}{}{%
    \nolinenumbers
}
\makeatother

\begin{document} 

\makeatletter
\def\linenumbers{\relax}
\def\thelinenumber{}
\makeatother
   \title{Modeling the formation of N$_2$ and CH$_4$ frost on the Pluto slopes}

   \author{L.Lange \inst{1}, T.Bertrand\inst{2,3}, V.Belissa \inst{3}, S.Capry \inst{3}, L.A. Young\inst{4}, A.Falco \inst{2,5} }

   \institute{NASA Postdoctoral Program Fellow at Jet Propulsion Laboratory (JPL), California Institute of Technology, Pasadena, CA 91011, USA.\\
              \email{lucas.lange@jpl.nasa.gov}
         \and
            Laboratoire d'Instrumentation et de Recherche en Astrophysique (LIRA) , Paris Observatory, Meudon, France, \and Nantes Université, Univ Angers, Le Mans Université, CNRS, Laboratoire de Planétologie et Géosciences (LPG),
LPG UMR 6112, 44000 Nantes, France.  
\and southwest Research Institute, Boulder, CO 80302, USA
\and Laboratoire de Météorologie Dynamique, Institut Pierre-Simon Laplace (LMD/IPSL), Sorbonne Université, Centre National de la Recherche Scientifique (CNRS), École Polytechnique, École Normale Supérieure (ENS), Paris, France.}

  \abstract
   {The equatorial region of Cthulhu as revealed by New Horizons appears to be generally dark and largely devoid of volatiles because its surface albedo is low. Localized bright patches, however, which are interpreted as CH$_4$ frost, are observed on crater rims and slopes. }
   {Previous studies suggested that these frosts might result from the peculiar insolation driven by the geometry of these slopes, but this has never been tested quantitatively. We investigated the origin, stability, and potential role of these localized frost deposits in the volatile cycle of Pluto.  
 }
   {We implemented a new subgrid-scale slope parameterization in the volatile transport model for Pluto, which accounts for the specific solar irradiation and the resulting surface and subsurface temperatures on sloped terrains. This parameterization also allows the condensation and sublimation of volatiles (either N$_2$ or CH$_4$) on slopes, including the effect of large-scale transport of these species. This is key to determining the amount of frost that forms or disappears. }
   {Our simulations reproduce the observed CH$_4$ frost on north-facing slopes as seasonal deposits that currently sublimate, predict perennial CH$_4$ frost on south-facing slopes, and show that the slope microclimates are not expected to alter  global volatile cycles.}
   {Seasonal and perennial N$_2$ and CH$_4$ frosts can form on the Pluto slopes, even in its darkest and warmest regions, because the locally sunlight received on inclined terrain is reduced. Despite the abundance of sloped surfaces on Pluto, the slope microclimates still only appear to have a minor effect on the global volatile cycles of the planet.}


   \maketitle

\section{Introduction}
The flyby of Pluto and its satellites by the NASA New Horizons spacecraft in July 2015 revealed a geologically and climatically dynamic world, with diverse surface ices and a tenuous but complex atmosphere \citep{Stern2015,Grundy2016,Gladstone2016,Moore2016}. The Multispectral Visible Imaging Camera (MVIC) and the Linear Etalon Imaging Spectral Array (LEISA) composing the Ralph spectrograph on board the spacecraft detected a spatially heterogeneous distribution of volatiles that are mainly composed of N$_2$, CH$_4$, and CO onto the water-ice bedrock \citep{Protopapa2017,Schmitt2017,Earle2018} that we summarize here. The most prominent N$_2$ ice deposit on the Pluto surface is the kilometer-thick sheet that is mixed with CH$_4$ and CO ice in Sputnik Planitia \citep{Stern2015,Grundy2016}. The rest of the planet exhibits a clear latitudinal stratification between CH$_4$-rich and N$_2$-rich terrains. The north pole is covered by relatively pure and bright CH$_4$-rich ice. The plains between 55\textdegree N–70\textdegree  N and 25\textdegree N–35\textdegree N  are covered in mixtures of CH$_4$ ice (possibly including CO and N$_2$ ice), with N$_2$ ice deposits filling the bottom of some craters and depressions. Between 35\textdegree N–55\textdegree N, the composition of the surface is dominated by N$_2$-rich ice deposits. The equatorial regions of Pluto (defined here as $\pm$15\textdegree~latitude) exhibit a strong longitudinal variability in the appearance and composition of the surface \citep{Buratti2017,Schmitt2017,Moore2018}, with the  N$_2$-ice rich deposit in Sputnik Planitia,  the relatively pure CH$_4$-rich ice deposits in the bladed terrain deposit, and the dark \citep[albedo of $\sim$ 0.1][]{Hofgartner2023} overall volatile-free plains in the Cthulhu region (defined here as the region between 20\textdegree N and 20\textdegree S, 0\textdegree E and 180\textdegree E, also known as the Belton region) . 

These observations, combined with numerical models such as volatile transport models \citep[e.g.,][]{Hansen1996,Young2012,Young2013,Bertrand2016,Bertrand2018,Bertrand2019}, or global climate models \citep[GCMs; e.g.,][]{Zalucha2013,Toigo2015,Forget2017}, show that the dynamics of these ice deposits are driven by latitude, elevation, seasonal evolution, and long-term orbital cycles \citep[see a review in][]{Moore2021,Young2021}. For example, \cite{Bertrand2016} and \cite{Bertrand2018} showed that the accumulation of N$_2$ ice occurs within Sputnik Planitia as a result of the lower elevation in the basin, which leads to a stronger thermal infrared cooling that is balanced by a higher condensation temperature of N$_2$. This in turn leads to a higher condensation rate than in any other location. \cite{Bertrand2019} showed that the orbital configurations of Pluto in the past Earth million years would have led to a net accumulation of CH$_4$-rich deposits in the equatorial regions of Pluto, and to seasonal accumulation elsewhere. The persistent absence of perennial ice in the equatorial region of Cthulhu may stem from its albedo, which is exceptionally low compared to the rest of the Pluto surface. This enhances solar heating of the high-inertia water-ice substrate. This stored heat is then released during colder seasons and keeps surface temperatures higher than elsewhere, which in turn inhibits CH$_4$ frost accumulation \citep{Earle2018alb}. In contrast, the high albedo of the seasonal CH$_4$ deposits at medium and high latitudes favors surface cooling in winter, promoting N$_2$ condensation and contributing to the formation of N$_2$-rich terrains at these latitudes \citep{Earle2018alb,Bertrand2019}. While these models successfully capture the first-order mechanisms controlling the large-scale ice distribution observed by New Horizons, it remains challenging to accurately quantify the amount of and sublimation rate or condensation because key surface and thermal parameters (e.g., the CH$_4$ ice albedo or the water-ice thermal inertia) are only loosely constrained by the available observations \citep{Young2021}.

Recent analyses using high-resolution measurements by Ralph/MVIC and the Long-Range Reconnaissance Imager (LORRI) instruments revealed bright N$_2$ and CH$_4$ ice-rich deposits on crater floors, rims, and mountain tops \citep{Bertrand2020,Earle2022}. More specifically,  \cite{Bertrand2020} highlighted the bright  CH$_4$ frost on the north-facing crater rims and slopes in the Cthulhu region and at the top of Pigafetta and Elcano Montes. Using high-resolution modeling, the authors showed that the methane deposits found at the top of these mountains form from a circulation-induced enrichment of gaseous methane a few kilometers above the Pluto plains that favors methane condensation at mountain summits. Their model cannot be used to study the formation of CH$_4$ deposits on crater rims, however, because the resolution of these craters is too thin compared to the model resolution. The authors suggested that these deposits found on the crater flanks might be linked to the specific microclimates of these slopes. Because the atmosphere of Pluto is tenuous, sloped terrains receive significantly different amounts of solar insolation compared to flat surfaces during  winter and spring (see their Figs. S8 and S9). This creates localized cold microclimates on the slope that can favor the accumulation of volatiles and mitigate their sublimation. They therefore suggested that the frosts observed on the Cthulhu crater rims are seasonal frost that formed during the previous winter and now sublimates. \cite{Earle2022} extended this work to other regions of Pluto and found a latitudinal dependence in the orientation of CH$_4$ frost on crater rims. They showed that methane frost is detected on north-facing slopes in the equatorial regions of Pluto ($\pm$ 15\textdegree~latitude), on equatorward-facing slopes at higher latitudes (between 15\textdegree~N and 38\textdegree N), and then on all of the crater rims at higher latitudes. Although they shared the hypothesis by \cite{Bertrand2020} to explain the methane frost at low latitudes on north-facing slopes, they noted that the change in preferential orientation in regions around 18\textdegree N might be explained by different formation timescales (e.g., seasonal at low latitudes, and perennial at high latitudes). Finally, for N$_2$ frosts, they showed that they are preferentially observed on the crater floor up to 38\textdegree~latitude and on equatorward-facing slopes at higher latitudes. 

No numerical model has been able to address the frost formation hypotheses proposed by \cite{Bertrand2020} and \cite{Earle2022} so far. These frosts and their associated slope microclimates occur on spatial scales of only a few kilometers \cite{Bertrand2020}, which is far smaller than the typical resolution of GCMs \citep[e.g., hundreds of kilometers,][]{Bertrand2019}. Local models, such as 1D radiative equilibrium models, are also inadequate because the stability of methane frost depends on its atmospheric mixing ratio, which is itself governed by large-scale meteorology that cannot be captured by these simplified models. This numerical gray zone has prevented further investigation into the dynamics of methane frosts on the Pluto slopes so far.

A similar problem was encountered by the Martian community with the modeling of water frost on tropical slopes. Its thin atmosphere is transparent to visible radiation, and slope-driven microclimates are therefore significant on Mars. These slope microclimates allow the formation of CO$_2$ frost (an analog for N$_2$ on Pluto) and water frost (an analog for CH$_4$ on Pluto) at subtropical latitudes, up to 20–30 degrees equatorward of the seasonal polar caps \citep[e.g.,][]{Vincendon2010water,Lange2024}. The same numerical limitations as described above applied when these low-latitude frosts on crater walls were modeled. Recently, \cite{Lange2023} introduced a subgrid parameterization of slope microclimates in the Mars Planetary Climate Model. The principle of this parameterization is illustrated in Fig.~\ref{fig:illustrationparam}. For each grid cell of the global Mars climate model, the terrain is decomposed into a distribution of sloped surfaces (defined by characteristic slope angles) and a flat surface. In each subgrid terrain, the surface energy balance and the condensation/sublimation of volatiles are computed under the assumption that all terrains interact with the same atmospheric column. The near-surface atmosphere then experiences an average effect of these subgrid microclimates. This parameterization successfully reproduced the seasonal dynamics of CO$_2$ and water frosts at low latitudes on Mars.

In this study, we adapted the slope microclimate parameterization developed by \cite{Lange2023} for Pluto to (1) describe the seasonal evolution of methane and nitrogen frost on crater slopes over a Pluto year, with particular attention to the sensitivity to thermophysical and optical properties of the ices, (2) discuss the nature of the frosts observed by New Horizons on crater flanks and slopes , and (3) discuss the potential effect of the slope microclimates on the global climate of Pluto. To this end, we describe the parameterization in Sect.~\ref{secmethods}. The dynamics of methane and nitrogen frost on sloped terrains and the nature of the frosts observed by New Horizons are presented in Sect.~\ref{sec:results}. The potential implications for the global climate of the slope microclimates are discussed in Sect.~\ref{sec:discussion}. We draw our conclusions in Sect.~\ref{sec:conclusions}.
  
\section{Model and methods \label{secmethods}}

\subsection{Frost identification and statistics}
\label{ssec:methodidentidification}
In order to compare our model with the observations of frost on slopes, we first determine the typical slope angle and azimuth at which frost is observed. While the mapping of N$_2$/CH$_4$ frosts has already been made in previous studies \citep[e.g.,][]{Schmitt2017,Earle2018}, we focus here on the Cthulhu region as it is an unique location where frost is only observed on slopes and not on the dark plains of the regions \citep{Bertrand2020}, suggesting the crucial role of slope microclimates -and thereby slope angle and azimuth- in the formation of these frosts. As in \cite{Bertrand2020}, we first map the location of bright deposits on LORRI observations \citep{Hofgartner2023}.  As some of the bright locations might be linked to direct illumination, we then compare the locations of these bright deposits to the CH$_4$ band depth map and phase index from the LEISA instrument \citep{Schmitt2017,Gabasova2021} to derive the nature of these bright surface -frost or not, and if so, if it is pure N$_2$, CH$_4$ diluted in N$_2$ ice or a CH$_4$-rich ice phases-. Finally, we determine the slope angle and azimuth at which ice is observed using the Digital Evolution Model (DEM, with a 300~m resolution per pixel) from \cite{Schenk2018}. While Hapke photometric analyses can, in principle, provide an independent estimate of surface roughness through the mean slope parameter ($\Bar{\theta}$), the restricted phase-angle range of New Horizons data leads to strong parameter degeneracies that prevent us from constraining $\Bar{\theta}$ \citep{Protopapa2020}. For this reason, and given that only localized applications of alternative photometric models exist \citep[e.g.,][]{Mishra2025}, we use the \cite{Schenk2018} DEM as it the most comprehensive and reliable dataset for slope measurements.

\subsection{The Pluto Volatile Transport Model \label{ssec:plutovtm}}
For this study, we use the Pluto Volatile Transport Model (VTM)  described by \citet[see their Methods section]{Bertrand2016}, which has been employed in previous investigations of ice dynamics on Pluto \citep{Bertrand2016,Bertrand2018,Bertrand2019}. In short, this model computes the surface radiative budget -assuming that the atmosphere is transparent at all wavelengths-, the exchange of heat with the subsurface by conduction, the exchange of heat with the surface by latent heat, and the exchanges of volatile (N$_2$ and CH$_4$) with the atmosphere. To represent horizontal transport after surface exchanges, the mixing ratios of atmospheric species are forced towards their global mean value, using a timescale of 10$^7$~s for CH$_4$, 1~s for N$_2$  estimated from 3D GCM studies \citep{Bertrand2016,Forget2017}.  

Simulations are performed with a horizontal grid with a grid-point spacing of 1\textdegree~by latitude and longitude ($\sim$~21~km at the equator), and a vertical grid dividing the atmosphere into 27 levels ranging from a few meters close to the surface up to 140~km. The initial distribution of the N$_2$ and CH$_4$ reservoirs is obtained following the 30 million years of volatile ice evolution performed by \cite{Bertrand2018} and \cite{Bertrand2019}: an infinite amount of N$_2$ ice is placed within Sputnik Planitia, and an infinite reservoir of CH$_4$ ice is placed in the Bladed Terrain Deposits (located between 15\textdegree S and 15\textdegree N, 140\textdegree W and 15\textdegree E). 

The emissivity and albedo of such ice deposits are adjustable parameters of the model, and  set to have an atmospheric surface pressure and CH$_4$ volume-mixing ratio (also called molar mixing ratio, that is,  the number of moles of the CH$_4$ gas divided by the number of moles of air, in percent) consistent with the measurements during the New Horizons fly-by (see Sect. \ref{ssec:inputs}). The albedo of the frost/ice-free terrains are set to observed values from \cite{Hofgartner2023} and their emissivity to 1. The diurnal thermal inertia of the bare-surface,  N$_2$ ($\rm{TI}_{\rm{N}_2,\rm{d}}$) and CH$_4$ ($\rm{TI}_{\rm{CH}_4,\rm{d}}$) are set by default to 9~J~m$^{-2}$~K$^{-1}$~s$^{-1/2}$ following low-thermal inertia derived by \cite{Lellouch2011} and \cite{Bertrand2025}. The seasonal thermal inertia of the bedrock $\rm{TI}_{\rm{bedrock},\rm{s}}$ is another adjustable parameter. The albedo of a surface is changed when ice/frost is present at the surface following the albedo-feedback parameterization of \cite{Bertrand2020}.

\subsection{Slope parameterization}

\subsubsection{Description of the parameterization for the Pluto VTM}
\label{ssec:slopeinVTM}
The slope parameterization in the Mars Planetary Climate Model proposed by \cite{Lange2023} relies on the assumption that the annual mean solar irradiance of a terrain with a slope angle $\theta$ varies linearly with the cosine of the slope azimuth $\psi$. Hence, they concluded given slope ($\theta$, $\psi$) can be thermally, on average, represented by a slope with a slope angle of $\mid \mu \mid $ that is either north-facing if $\mu~>~0$, or south-facing if $\mu~<~0$, where $\mu$ is the projected slope in the meridional direction,

\begin{equation}
\mu = \theta \cos(\psi)  
\label{eq:defmu}.
\end{equation}
This conclusion must be qualified for east-west oriented slopes in equatorial regions on Mars, however. The linear correlation observed between annual-mean solar irradiance on a slope and the projected slope  $\mu$ at high latitudes is primarily due to the preferential azimuthal position of the Sun throughout the Martian year. At lower latitudes (i.e., within the tropics), the obliquity of Mars (25.2\textdegree) causes the solar declination to alternate between the northern and southern celestial hemispheres. This seasonal symmetry disrupts the dominance of a single solar azimuth and reduces the strength of the correlation between annual irradiance and $\mu$. Solar irradiance calculations for 30\textdegree~slopes at various latitudes show that the correlation with $\mu$ holds only above approximately 10\textdegree~latitude, with the values of the determination coefficient, R$^2$,  dropping below 0.9 closer to the equator. Hence, the approximation proposed by \cite{Lange2023} that the annual thermal behavior of a given slope ($\theta,\psi$) can be represented by its projected slope $\mu$ remains valid for present-day Mars at latitudes outside the equatorial band.

\begin{figure}
   \centering
   \includegraphics[width=0.4\textwidth]{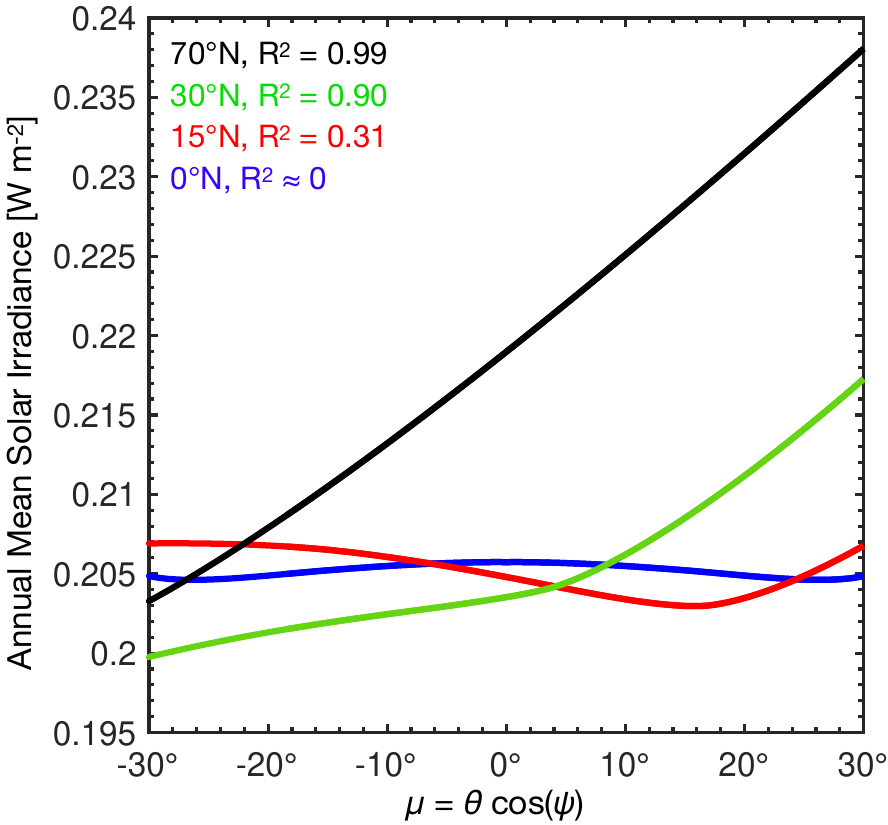}
   \caption{ Annual mean solar irradiance (at the surface) for several slope angles $\theta$ and azimuths $\psi$ ($\psi$~=~0\textdegree~, corresponding to a slope oriented northward, at a latitude of 70\textdegree N (black), 30\textdegree N (green), 15\textdegree N (red), and 0\textdegree~(blue). The correlation between the annual mean solar irradiance  and the projected slope $\mu$ for each latitude is given by the determination coefficient R$^2$. }
              \label{fig:insolation1DPluto}
              
    \end{figure}

The latter conclusion cannot be directly applied to Pluto. Given its high obliquity -present-day value of 119.6\textdegree-, and thereby the high latitudes of the tropics ($\sim$~60\textdegree), the correlation between the yearly-solar irradiance and slope azimuth is not straightforward on most of the Pluto surface. When computing the solar irradiance for several steep slopes (Fig. \ref{fig:insolation1DPluto}), we observe that the correlation is only clear for high latitudes ($\ge$~70\textdegree), and disappears then rapidly at lower latitudes.  Below this threshold, slopes with a significant east–west component cannot be thermally approximated by their projected slope $\mu$.

\begin{figure*}[h!]
   \centering
   \includegraphics[width=\textwidth]{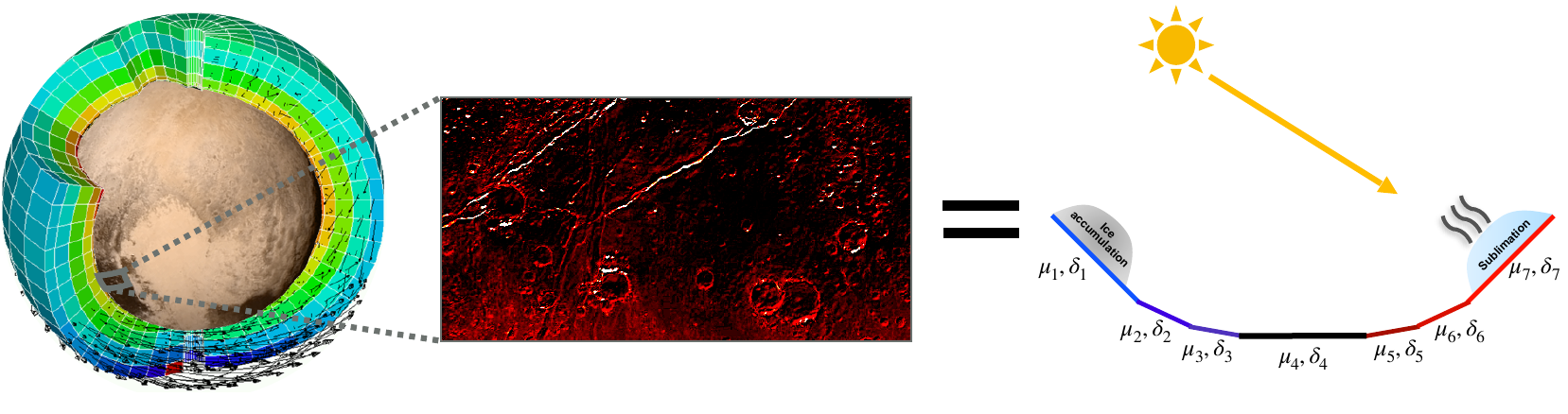}
   \caption{Schema of the subgrid-scale slope parameterization. Each coarse mesh of the global Pluto Volatile Transport Model/Pluto Planetary Climate Model (left) is decomposed into north-south facing subgrid slopes (defined by characteristic slopes $\mu_i$; right) or flat terrain. Each of these subgrid surfaces is associated with its cover fraction $\delta_i$, which represents the percentage of the mesh grid occupied by a given slope  $\mu_i$ within the observed local topography (middle). These subgrid terrains have their own microclimate (insolation, surface, subsurface temperatures, condensation/sublimation of frosts, etc.), and the interactions between the atmosphere and surface are made through averaged values over the mesh. The bluish surfaces on the right show slopes that are colder than the flat surfaces (and warmer for reddish surfaces).}
              \label{fig:illustrationparam}
              
    \end{figure*}

Based on these considerations, our parameterization is restricted to the evolution of CH$_4$ and N$_2$ frost on north–south-oriented slopes, as slopes with an east–west component are likely not accurately represented. To implement this, each grid cell of the Pluto VTM is divided into a set of subgrid units representing either north-facing slopes, south-facing slopes, or flat terrain (Fig.~\ref{fig:illustrationparam}). This simplification is justified by 1) the preferential accumulation of CH$_4$ frost on pole-facing slopes at low latitudes \citep[][ Sect.~\ref{ssec:resultfrostobs}]{Earle2022}, and 2) the fact that explicitly resolving all slope orientations would require introducing a larger number of subgrid-slope categories, leading to a prohibitive increase in computational cost.

\begin{figure}[h!]
   \centering
   \includegraphics[width=0.45\textwidth]{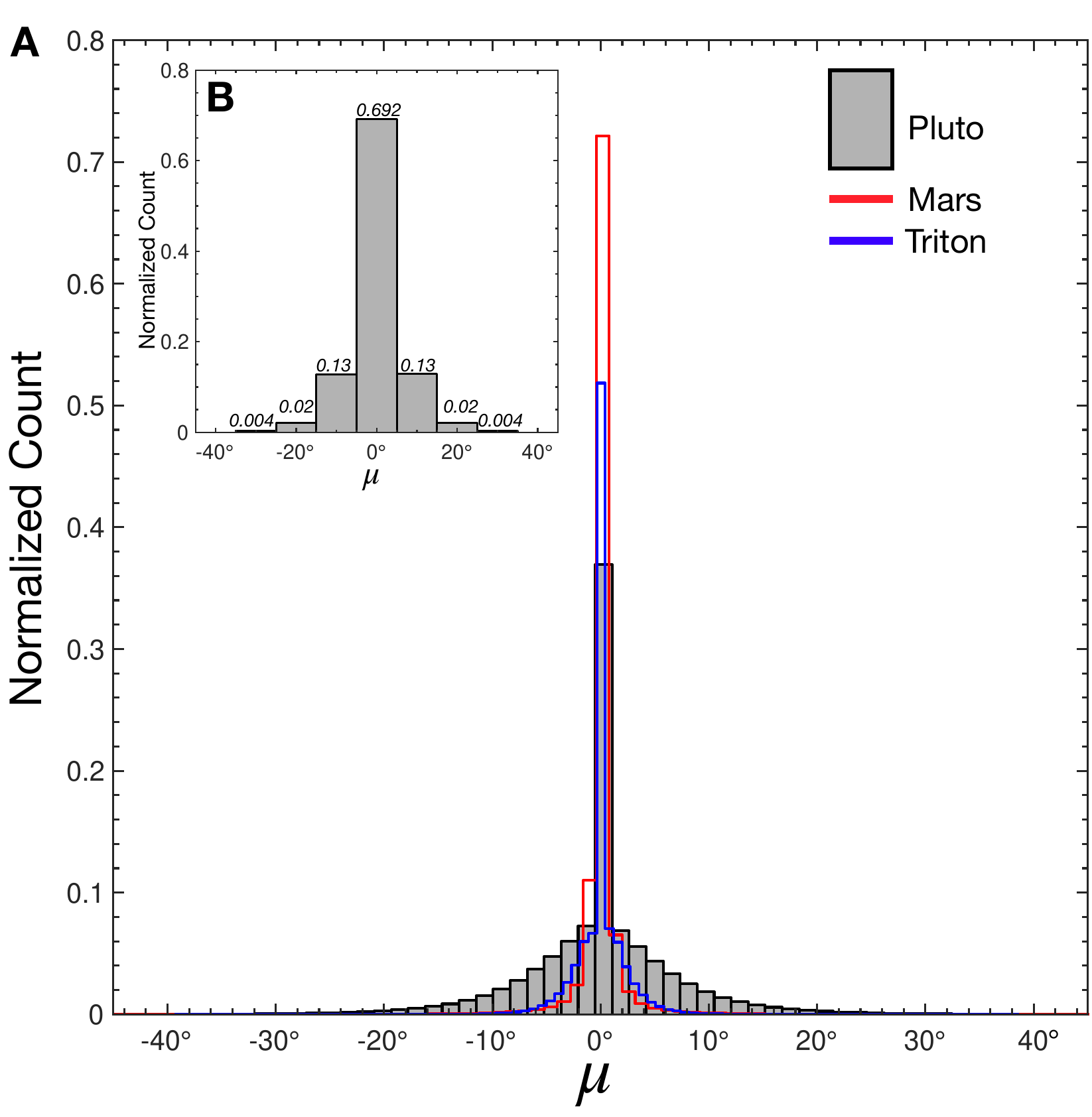}
   \caption{(A): Normalized distribution of projected slopes $\mu$, where $\mu$~=~0\textdegree~corresponds to flat and east-west-facing slopes, on Pluto (solid gray histogram), Mars \citep[red, from][]{Lange2023}, and Triton \citep[derived from the fractional data  of][]{Schenk2021}.
(B): Cover fraction $\delta_i$ derived from panel A), obtained by rebinning the histogram into the $\mu_i$ bins defined in the main text, each with a width of $\pm$ 5\textdegree. }
              \label{fig:histmu}
              
    \end{figure}

 We use seven subgrid slopes, three north-facing, three south-facing, and one flat surface, with $\mid{\mu_1}\mid~=~\mid{\mu_7}\mid$~=~30\textdegree, $\mid{\mu_2}\mid~=~\mid{\mu_6}\mid$~=~20\textdegree, $\mid{\mu_3}\mid~=~\mid{\mu_5}\mid$~=~10\textdegree~and $\mu_4$~=~0\textdegree. The number of seven subgrid slopes is chosen from our Mars'experience \citep{Lange2023}, which showed that such values allow a correct representation of the distribution of slopes within a mesh, without increasing the computational time significantly. Within each grid cell, every subgrid slope is assigned a fractional coverage $\delta_i$, representing the percentage of the cell area occupied by slopes whose projected inclination lies within $\mu_i \pm 5$\textdegree. The cover fraction associated with a given projected slope angle $\mu_i$ is computed as follows. At each point of the Pluto DEM, we calculate the local slope and azimuth, and derive the corresponding $\mu$ value using Eq.\ref{eq:defmu}. When the slope is predominantly oriented east–west (defined as an azimuth $\psi$ within $\pm$45\textdegree~of either 90\textdegree~or 270\textdegree), we assume it behaves similarly to a flat surface and assign $\mu = 0$\textdegree. The resulting global distribution of $\mu$ is shown in Fig.\ref{fig:histmu}a.  From this distribution, we define the cover fraction $\delta_i$ as the proportion of terrain with $\mu$ values within $\mu_i \pm 5$\textdegree, normalized by the total number of samples (Fig.~\ref{fig:histmu}b). This distribution is then applied uniformly to all grid cells in the VTM, regardless of geographic location. We adopt this approach to avoid introducing artificial biases that could arise from the incomplete coverage of the DEM, particularly in the southern hemisphere, where no topographic data are available. Attempting to extrapolate or impose subgrid-slope distributions in these poorly constrained regions could lead to spurious volatile fluxes in the model, for instance, by artificially channeling volatiles from smooth to rougher mesh cells. While it is likely that some regions of Pluto have indeed more slopes than others, using a single, globally averaged distribution ensures that such modeling artifacts are avoided and that simulations remain internally consistent despite the limitations of the topographic dataset. Moreover, because the VTM does not resolve atmospheric circulation and instead applies a uniform relaxation toward equilibrium, volatile transport remains simplified . This simplification means that potential transfers from rough to smooth terrains are not explicitly captured, which could otherwise bias flux estimates. We therefore retain a single, global slope distribution, and our results should be interpreted as showing where N$_2$ or CH$_4$ frost can form on a given slope rather than providing an accurate quantification of lateral volatile fluxes or total mass budgets for the frosts/ice.

The slope surface temperature is computed based on its energy budget given by

\begin{equation}
\begin{split}
\rho  c_s \frac{\partial T_{\rm{surf,slope}}}{\partial t} &= (1-A_{\rm{slope}}) F_{\rm{rad,sw}}(t) + \epsilon_{\rm{slope}}  F_{\rm{rad,lw}}(t) +  F_{\rm{ground}}(t) \\
  &\quad + \sum_{i}L_i\frac{\partial m_i}{\partial t}(t)   - \epsilon_{\rm{slope}} \sigma T_{\rm{surf,slope}}^4(t)
\end{split},
\label{Eq:energysurflayer}
\end{equation}

\noindent where the left member of the equation is the energy of the surface layer, with $\rho$ is the  density of the ground (kg~m$^{-3}$),   $c_s$ is the surface layer heat capacity per unit area (J~m~kg$^{-1}$~K$^{-1}$),  $T_{\rm{surf,slope}}$ the subgrid-slope temperature (K), $A_{\rm{slope}}$ (unitless) the albedo of the surface, $F_{\rm{rad,sw}}$ the radiative flux at visible wavelengths (W~m$^{-2}$), $\epsilon_{\rm{slope}}$~(unitless) is the surface emissivity (assumed to be equal to the absorptivity), $F_{\rm{rad,lw}}$ the radiative flux at infrared  wavelengths (W~m$^{-2}$), $F_{\rm{ground}}$ the soil heat flux due to heat conduction process (W~m$^{-2}$), $\sum_{i}L_i\frac{\partial m_i}{\partial t} $ the latent heat flux due to the condensation/sublimation of a volatile with a latent heat $L_i$ (J~kg$^{-1}$) and $ \epsilon_{\rm{slope}} \sigma T_{\rm{surf,slope}}^4$ the radiative cooling of the surface with $\sigma$~=~5.67$\times 10^{-8}$ W~m$^{-2}$~K$^{-4}$ the Stefan-Boltzmann constant.

Assuming that  the atmosphere is transparent at all wavelengths, and allowing only reflection of the solar radiation between the slope and the surrounding plains,  the total solar irradiance on a given subgrid slope $F_{\rm{rad,sw}}$ is given by
\begin{equation}
    F_{\rm{rad,sw}}(t) = \frac{S_0}{R(t)^2}\left(\sin(\theta_{\rm{slope}}(t)) + A_{\rm{flat}}(t) \sigma_{s} \sin(\theta_{\rm{flat}}(t)) \right),
    \label{eq:Fradsw}
\end{equation}

\noindent where $S_0$~=~1361~W~m$^{-2}$ is the solar constant, $R$ is the distance between the Sun and Pluto (in astronomical units), $\theta_{\rm{slope}}$  is the angle of the Sun above the sloped surface (when  $\sin(\theta_{\rm{slope}}(t)) < 0$, i.e., when the Sun if below the horizon, or the slope is in a self-shadow, then we set $\sin(\theta_{\rm{slope}}(t))$ to 0); and $\theta_{\rm{flat}}$ is the same angle but for the flat surface, $A_{\rm{flat}}$ is the albedo of the flat surface, and $\sigma_s$ (unitless) is the  sky-view factor, which  quantifies the proportion of the sky in the half hemisphere seen by the slope that is not obstructed by the surrounding terrain, computed with
\begin{equation}
\sigma_s    = \frac{1+\cos(\mid \mu \mid )}{2},
\end{equation}
\noindent where $\mid \mu \mid$ refers to the absolute value of the projected slope $\mu$.  

Assuming again that the atmosphere is transparent, $F_{\rm{rad,lw}}(t)$ is then just computed with

\begin{equation}
    F_{\rm{rad,lw}}(t) = (1-\sigma_s)\epsilon_{\rm{flat}}(t)\sigma T_{\rm{surf,flat}}(t)^4
    \label{eq:Fradlw}.
\end{equation}

The computation of the ground flux $F_{\rm{ground}}$ is based on the  1D soil model originally described by \cite{Hourdin1993}. To adequately resolve the diurnal/seasonal scale lengths on Pluto, the subsurface is divided into 24 discrete layers, with a geometrically stretched distribution, with the shallowest layer at 1.4~10$^{-4}$~m
and the deepest layer depth at nearly 1000~m.

The amount of N$_2$ frost condensing/subliming on a slope $\frac{\partial m_{\rm{N}_2}}{\partial t}$~(kg~m$^{-2}$)  is derived from the amount of latent heat required to keep temperatures at the frost point if N$_2$ ice is present, that is,
\begin{equation}
   \frac{\partial m_{\rm{N}_2}}{\partial t}(t) = \frac{\rho c_s}{\Delta t L_{\rm{N}_2}} \left(T_{\rm{cond}} - T_0 \right),
   \label{eq:TcondN$_2$_nogcm}
\end{equation}
\noindent where $\Delta t$~(s) is the timestep of the model, $L_{\rm{N}_2}$~=~2.5~10$^5$~J~kg$^{-1}$ is the latent heat of condensation for nitrogen, $T_{\rm{cond}}$ nitrogen condensation temperature derived from the thermodynamic relations of \cite{Fray2009}, and $T_0$ is the surface temperature on a subgrid slope predicted by radiative and conductive balance $T_0$ (i.e., Eq. \ref{Eq:energysurflayer} solved without any condensation/sublimation of N$_2$). This term is set to zero when no ice is present and $T_0 > T_{\rm{cond}}$. If the amount of ice subliming is too high compared to what is available at the surface, then  $- \frac{\partial m_{\rm{N}_2}}{\partial t}(t)$ is  set to the amount of ice available divided by the timestep.

The amount of CH$_4$ frost condensing/subliming on a slope is given by
\begin{equation}
   \frac{\partial m_{\rm{CH}_4}}{\partial t}(t) = \rho C_d U \left(q - q_{\rm{surf}} \right)
   \label{eq:dmdtCH$_4$},
\end{equation}
\noindent where $\rho$~(kg~m$^{-3}$) is the near-surface air density, $C_d$~(unitless) is a near-surface drag coefficient set to 0.06 \citep{Forget2017}, $U$ is the wind velocity near the surface and is set to 0.5 m~s$^{-1}$ (based on 3D GCM studies), $q$~(in kg~kg$^{-1}$) the CH$_4$ atmospheric mass mixing ratio near the surface, and  $q_{\rm{surf}}$~(in kg~kg$^{-1}$) is the saturation vapor pressure mass mixing ratio at the surface, computed using the thermodynamic Claudius–Clapeyron relation of \cite{Fray2009} for CH$_4$ with a latent heat for sublimation $L_{\rm{CH}_4}$~=~586.7 kJ~kg$^{-1}$. Again, if the latter equation predicts CH$_4$ ice sublimation and no ice at the surface, this term is set to zero. 

Equation~\ref{Eq:energysurflayer} is solved using an implicit scheme. It is nontrivial to apply this scheme, however, because the surface temperature is intrinsically coupled to subsurface heat conduction, which links the surface radiative budget to the thermal diffusion within the ground \cite[see for instance the mathematical derivations by][]{Young2012}. To address this, we adopt the strategy proposed by \citep{Hourdin1995}: the internal energy of the surface layer (and thus the surface temperature) is first updated at time $t$ using Eq.~\ref{Eq:energysurflayer}. The temperatures of the subsurface layers are then computed by solving the heat conduction equation. Once these temperatures are known, the resulting soil heat flux is calculated and used as an input to Eq.~\ref{Eq:energysurflayer} at the next time step. This iterative procedure yields a stable and fully implicit scheme for our model.

Once each subgrid-slope variables (e.g., slope temperature, and albedo) are computed, grid parameters are determined by averaging all subgrid parameter values weighted by their respective cover fractions. The grid box albedo and emissivity can be computed with 
\begin{equation}
     X = \sum_{i=1}^7 X_i \delta_i
     \label{eq:opticalproperties},
\end{equation}
\noindent where $X$ refers to the grid parameter, $X_i$ the subgrid parameter of slope $i$, $\delta_i$ the cover fraction.

For the amount of volatile in the grid-box mesh, one must consider that the amount of ice on a subgrid surface is expressed as a surface mass density (kg~m$^{-2}$). The area of the sloped subgrid surface can be higher if it is sloped or flat, however, when a similar latitude/longitude boundary is assumed for the cell. Hence, to ensure mass conservation, the grid-box  amount of volatile of a species $X$ in is computed with 
\begin{equation}
     X = \sum_{i=1}^7 X_i \frac{\delta_i}{\cos(\mu_i)}
     \label{eq:traceurs}.
\end{equation}
Finally, the  grid-box surface temperature is derived by averaging the Stefan–Boltzmann functions

\begin{equation}
    \epsilon \sigma T_{\rm{surf,grid}}^4 = \sum_{i=1}^7 \epsilon_i \sigma T_{\rm{surf,i}}^4 \delta_i
    \label{eq:Tsurf},
\end{equation}

\noindent where $\epsilon$~(unitless) is the grid emissivity, $T_{\rm{surf,grid}}$~(K) the grid surface temperature, and quantities with subscript $i$ refer to the same quantities but for the subgrid slope.

\subsubsection{Modifications of the parameterization for the planetary climate model for Pluto}
The parameterization developed here for the Pluto VTM can be easily applied to any Pluto GCMs \citep[e.g., the Pluto Planetary Climate Model,][]{Forget2017} by making the following modifications: 

\begin{enumerate}
    \item The visible fluxes should be computed considering the effects of aerosols/hazes in the atmosphere. To do so, one can compute this solar irradiance on a flat surface, as traditionally done by radiative transfer algorithms \citep[e.g.,][]{Forget2017,Bertrand2017} and then derive it for any slope based on geometric considerations \citep[see for instance Eqs. 3 and 4 in][]{Spiga2008}. One should also estimate the scattered irradiance on slopes by the aerosols. This requires constructing a complete parameterization using radiative transfer models and is beyond the scope of this paper. We leave this for future work.
    \item The radiative effect of the atmosphere at infrared wavelengths should also be included in the surface energy budget. To do so, one can derive the downwelling infrared radiance $F_{\rm{rad,lw, flat}}$  for a flat surface using radiative transfer models \citep[e.g.,][]{Forget2017,Bertrand2017}, and then add the term $\sigma_s F_{\rm{rad,lw, flat}}$  in Eq. \ref{eq:Fradlw}.
        \item Near surface winds $U$ in Eq. \ref{eq:dmdtCH$_4$} should not be constant but derived by the Global Circulation Model. 
        \item When N$_2$ is condensing, Eq. \ref{eq:TcondN$_2$_nogcm} should be replaced by \citep{Forget2017}

\begin{equation}
   \frac{\partial m_{\rm{N}_2}}{\partial t}(t) = \frac{\rho c_s}{\Delta t} \frac{ \left(T_{\rm{cond}} - T_0 \right)}{L_{\rm{N}_2}+c_p \left(T_1 - T_{\rm{cond}}  \right)}
   \label{eq:TcondN$_2$_nogcmbis},
\end{equation}
\noindent where $c_p$ is the air specific heat at constant pressure (set here to 1040~J~kg$^{-1}$~K$^{-1}$), $T_1$~(K) the atmospheric temperature in the first layer of the model (typically a few meters). The additional terms $c_p \left(T_1 - T_{\rm{cond}}  \right)$ correspond to the heat brought by the atmosphere when cooled to the condensation temperature of N$_2$. When N$_2$ ice is subliming, Eq. \ref{eq:TcondN$_2$_nogcm} is used.
\end{enumerate}

\subsubsection{Validation and limits of the model}
\label{ssec:limits}
Validating the parameterization on Pluto is challenging due to the paucity of measurements, especially at the slope scale, and the strong sensitivity of ice dynamics to surface properties as highlighted by previous studies \citep[e.g.,][]{Bertrand2018,Bertrand2019}. Given the extensive validation of the parameterization on Mars using various datasets performed by  \cite{Lange2023}, we will consider that the application of the parameterization on Pluto is reliable. Yet, our parameterization has several limits that we acknowledge here. First, as mentioned previously, the microclimate of  east-west-facing slopes is not modeled here.  Second, shadows cast by surrounding terrain (as opposed to slope-induced self-shadowing) are neglected. This omission prevents the representation of cold traps, for example, at the bottom of a ridge/canyon. Modeling such shadowing would require ray-tracing computation for each point modeled here at each time of the year, leading to a too significant increase in the computation time.  Third, slope winds are not explicitly modeled: wind speeds on subgrid slopes are assumed to follow large-scale winds simulated in the Pluto GCM or prescribed in the VTM. Katabatic winds might induce stronger winds on crater rims, however, as was modeled on the flanks of the Sputnik Planitia basin \citep{Forget2017,Bertrand2020}, for instance. These slope winds should impact the sublimation rate of methane deposits as shown by Eq. \ref{eq:dmdtCH$_4$}. In the absence of slope-wind parameterization in our model, we test the sensitivity of our results to these slope winds by varying $U$ in  Eq. \ref{eq:dmdtCH$_4$} (see Table \ref{table:inputs}). Finally, a shared atmospheric column is assumed for all subgrid slopes, implying effective mixing by large-scale or regional circulation. Testing this would require mesoscale or LES modeling, beyond the scope of this study.

\subsection{Description of the simulations \label{ssec:inputs}}
As mentioned in Sect. \ref{ssec:plutovtm}, the optical properties of N$_2$ and CH$_4$ ices, and the thermal inertia of the substrate must be adjusted to reproduce the surface pressure and CH$_4$ volume-mixing ratio observed by stellar occultations and high-resolution spectroscopy respectively \citep[e.g.,][]{Hinson2017, Meza2019,Lellouch2015}. The default parameters listed in Table \ref{table:inputs} provide a good representation of the seasonal evolution of surface pressure and CH$_4$ volume-mixing ratio, in agreement with the observations of \cite{Hinson2017}, \cite{Meza2019}, and \citet[see, for instance, Fig. 2 of \citealt{Bertrand2016}]{Lellouch2015}. Given the sensitivity of the model to the inputs listed in Table \ref{table:inputs}, we investigate the robustness of our results by varying the input parameters. Because modifying the albedo or emissivity of the main ice reservoirs on Pluto would affect the global surface pressure and CH$_4$ levels, which in turn would affect the frost condensation/sublimation on slopes, we chose to vary the surface and ice properties in our sensitivity only within the longitudes of the Cthulhu region, between latitudes 40\textdegree N and 40\textdegree S. A complete description of the sensitivity of the CH$_4$ and N$_2$ cycles to these adjustable parameters can be found in \cite{Bertrand2016,Bertrand2018} and \cite{Bertrand2019}.

The baseline values were retained for the rest of the surface to maintain realistic pressure and CH$_4$ cycles.

\begin{table*}[h!]
\caption{Input parameters for the Pluto VTM and the values adopted in our sensitivity study.}
\begin{tabular}{c c c}       
\hline                
Parameter & Baseline Value & Other Values explored (in Cthulhu only)   \\    
\hline  \hline                      
   N$_2$ ice - albedo & 0.8 & 0.78-0.82  \\     
   N$_2$ ice - emissivity & 0.9 & 0.88-0.92  \\
   CH$_4$ ice at high-latitudes - albedo & 0.71 & /  \\
   CH$_4$ ice in the Bladed Terrain - albedo & 0.55 & /  \\
   CH$_4$ ice in the Bladed Terrain and at high-latitudes - emissivity & 0.8 & /  \\
   CH$_4$ ice in  Cthulhu - albedo & 0.67 & 0.67-0.75  \\
   CH$_4$ ice in  Cthulhu - emissivity & 0.8 & 0.79-0.81  \\
   $\rm{TI}_{\rm{CH}_4,\rm{d}}$ & 9 & 20 \\
   $\rm{TI}_{\rm{CH}_4,\rm{s}}$ & 800 & 600, 1000 \\
   $\rm{TI}_{\rm{N}_2,\rm{d}}$ & 9 & 5, 20 \\
   $\rm{TI}_{\rm{N}_2,\rm{s}}$ & 800 & 600, 1000 \\
   $\rm{TI}_{\rm{bedrock,~ s}}$ & 800 & 600, 1000 \\
   Large-scale winds $U$ & 1 & 0.1, 2 \\

\hline                                   
\end{tabular}
\tablefoot{For the sensitivity study, the parameter values were only varied within the Cthulhu region, while elsewhere, they remained fixed at their baseline values. The transition between the albedo of the bare surface and that of a CH$_4$-frosted surface follows the parameterization of \cite{Bertrand2020}. The values listed in this table correspond to the maximum albedo that a frosted surface can reach, typically for deposits thicker than $\sim$~1~mm.  TI refers to the thermal inertia, the subscript d corresponds to the diurnal thermal inertia, i.e., the first  centimeters of the surface, and the subscript s refers to the seasonal thermal inertia (depths deeper than $\sim$~5~cm). The baseline values for the first five rows were adjusted to ensure a realistic representation of the seasonal N$_2$ and CH$_4$ cycles.}             
\label{table:inputs}     
\end{table*}

For each set of parameters, the model was run for 100 Pluto years to equilibrate the subsurface temperatures on each subgrid slope, with slope condensation/sublimation disabled to avoid the formation or disappearance of thick glaciers due to non-equilibrated temperatures. We then ran the model for an additional ten years, allowing the condensation of volatiles on each subgrid slope. Only results from the final year are presented in this manuscript.

\section{Results \label{sec:results}}
\subsection{Frost mapping and statistics \label{ssec:resultfrostobs}}

  \begin{figure}[h!]
   \centering
   \includegraphics[width=.5\textwidth]{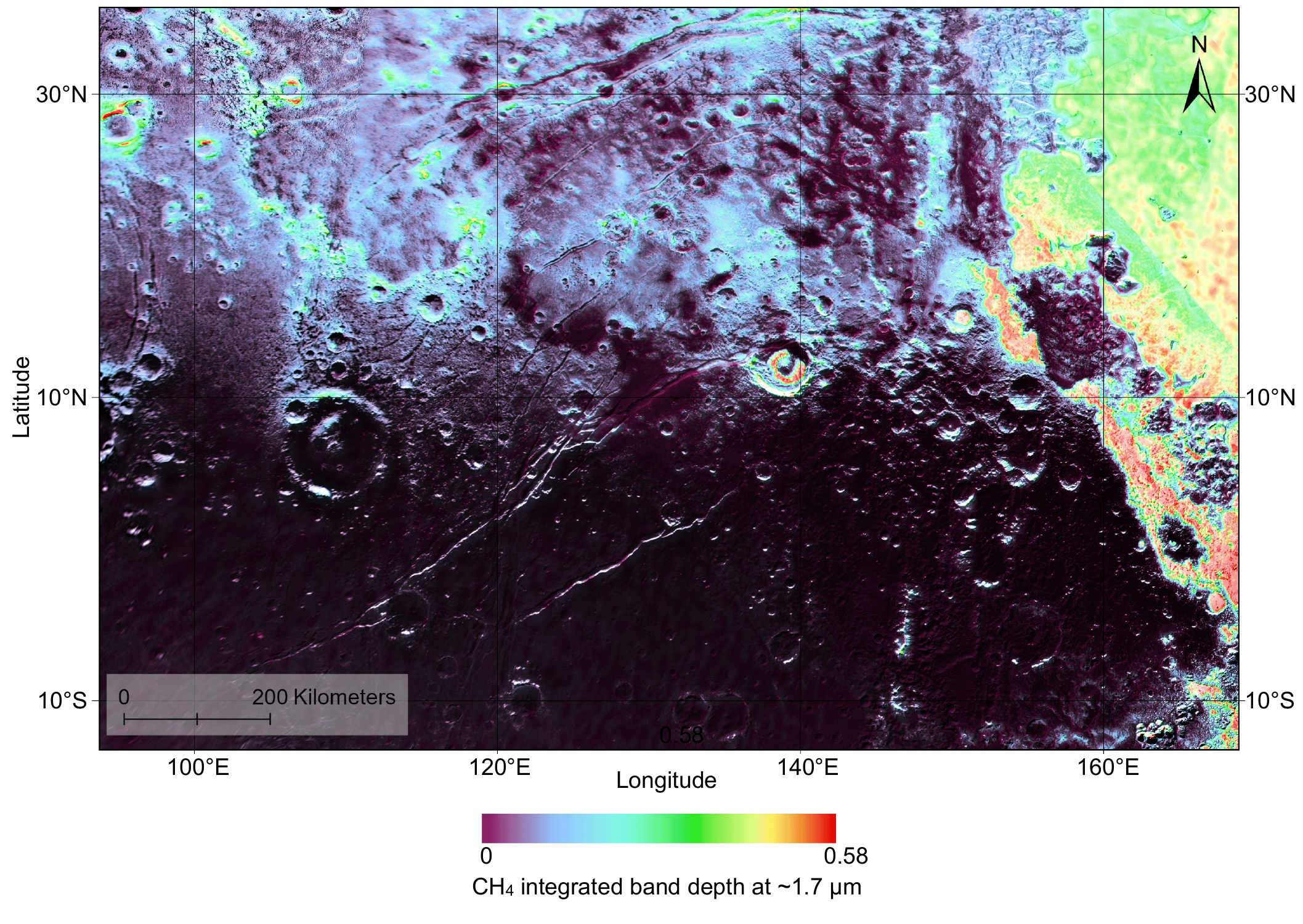}
   \caption{CH$_4$-integrated LEISA band depth maps from \cite{Schmitt2017} and \cite{Gabasova2021}, overlaid with the LORRI albedo map \citep{Hofgartner2023}. The white patches in the  Cthulhu region, which are not colored, are considered to be CH$_4$ frost (see the main text for an explanation).}
              \label{fig:mapfrost}
              
\end{figure}
The geographic distribution of frost identified in this study, based on the methodology described in Sect. \ref{ssec:methodidentidification} is presented in Fig.~\ref{fig:mapfrost}. A comparison between the LEISA CH$_4$ and N$_2$ band depth maps at this location indicates that these frosts correspond to N$_2$ ice deposits in Sputnik Planitia (northeastern part of Fig.~\ref{fig:mapfrost}) and Elliot crater (located around ~12\textdegree N, 139\textdegree E), and to CH$_4$ frost elsewhere. CH$_4$ frost is detected poleward of 10\textdegree N, but its identification at lower latitudes is more difficult, as the bright spots observed with LORRI are not always associated with a distinct signal in the CH$_4$ integrated LEISA band depth. This bias arises from the resolution difference between the LORRI observations \citep[$\sim$~300~m per pixel,][]{Weaver2008,Hofgartner2023} and the LEISA CH$_4$ band depth maps \citep[$\sim$~2.7~km at best,][]{Gabasova2021}.

Nonetheless, several lines of evidence support the interpretation that these bright deposits are indeed CH$_4$ frost. First, in the largest bright patches, a few pixels exhibit high CH$_4$ integrated band depth values (up to 0.4). In such cases, the correlation between bright deposits and CH$_4$ frost is robust, suggesting that the absence of a correlation elsewhere is primarily due to resolution differences between the instruments. Second, \cite{Earle2018} identified CH$_4$ ice deposits on slopes between 15\textdegree S and 15\textdegree N using MVIC data, which offers higher spatial resolution \citep[$\sim$~1~km][]{Earle2022}. Third, analysis of RGB images at these locations reveals that some bright, high-albedo slopes \citep[Supplementary Fig. 6 of][]{Bertrand2020} are adjacent to similarly oriented, red, volatile-free slopes \citep{Earle2022}, indicating that the bright signatures are not solely due to illumination effects, but also to the presence of surface ice. We therefore conclude that the bright deposits observed in Fig.~\ref{fig:mapfrost} within the Cthulhu region are CH$_4$ frost. The LEISA phase-index map \citep[see Supplementary Fig. 7b of][]{Bertrand2020} suggests that these deposits are relatively pure CH$_4$. \cite{Earle2022} further suggested that pure CH$_4$ ice is observed on crater rims at low latitudes, while CH$_4$ diluted with N$_2$ may occur on the north-facing crater walls at lower elevations.

\begin{figure}[h!]
   \centering
   \includegraphics[width=0.45\textwidth]{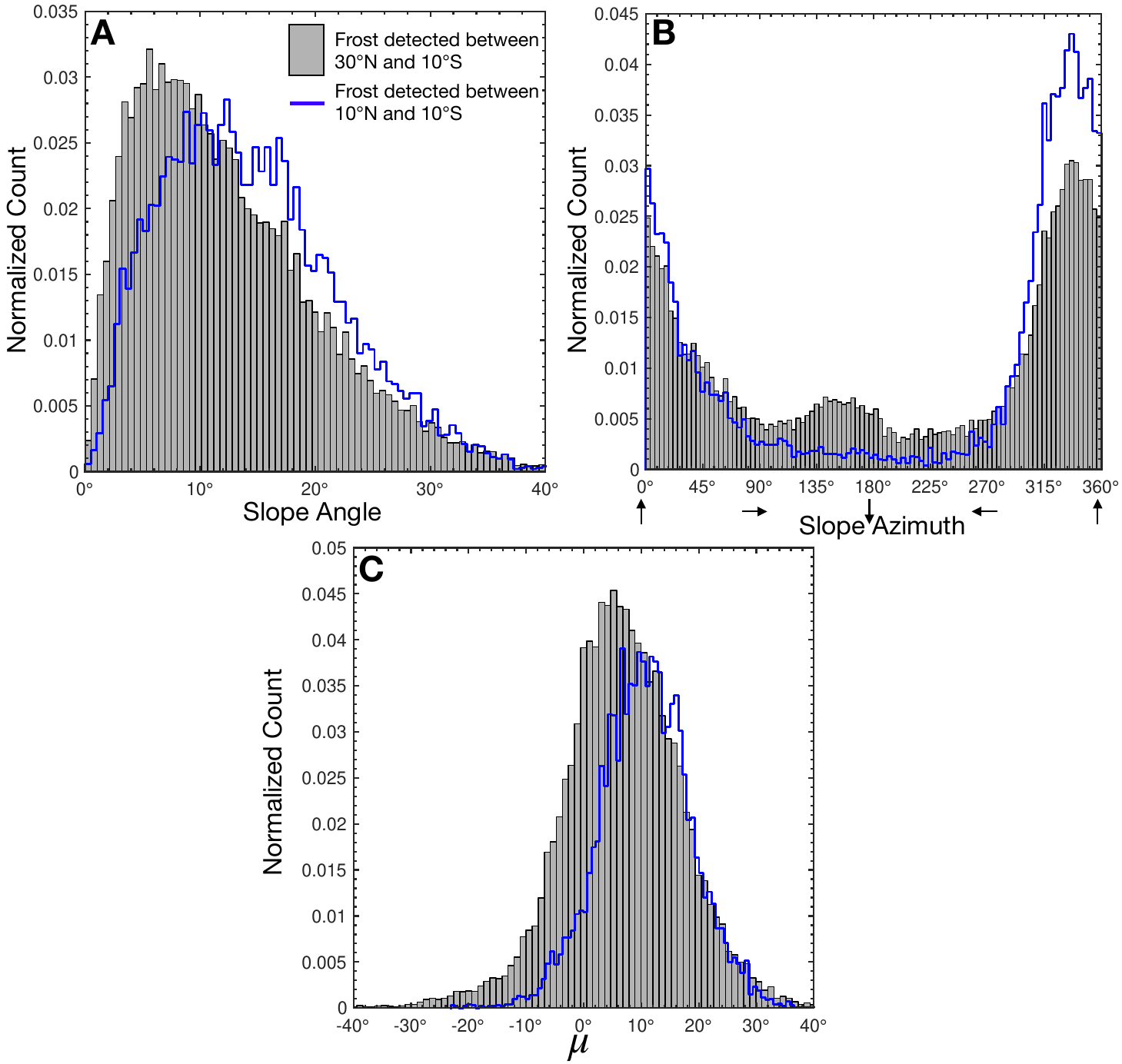}
   \caption{Normalized count of the slope angle (A), the slope azimuth (B), and the projected slope $\mu$ (C) where CH$_4$ frost is detected for all locations in Fig. \ref{fig:mapfrost} (gray histogram) and the Cthulhu region alone ($\pm$ 10\textdegree~latitudes; blue histogram).  }
              \label{fig:frostdetectedstat}
              
    \end{figure}

The slope angle, slope azimuth, and projected slope $\mu$ of the slopes where CH$_4$ frost is detected are shown in Fig.~\ref{fig:frostdetectedstat}. Note that this distribution, and the statistics derived below, are based on the observational dataset obtained by combining the \cite{Schenk2018} DEM with CH$_4$ frost detections (as described in Sect.~\ref{ssec:methodidentidification}), and are therefore not related to the discrete slope values or slope distribution used in our model simulations (Sect.~\ref{ssec:slopeinVTM}). The frost shown in Fig.~\ref{fig:mapfrost} occurs on both steep slopes (up to 53\textdegree) and flat terrain, particularly poleward of 10\textdegree N. The median slope angle at which frost is detected is 10.9\textdegree~$\pm$~8\textdegree~(1$\sigma$), with a clear preference for north-facing slopes, although a small number of south-facing slopes is also identified (Fig.~\ref{fig:frostdetectedstat}). north of 30\textdegree N, CH$_4$ frost preferentially forms on slopes facing south rather than north, a latitudinal transition approximately 15\textdegree~higher than that reported by \cite{Earle2022}. Although we also detect frost on slopes facing south between 15\textdegree N and 30\textdegree N, the dominant orientation remains north. We interpret this result with caution, however, because the low illumination on these slopes, especially at low latitudes, may hinder frost detection by reducing the signal-to-noise ratio of the spectroscopic data. In the specific case of the Cthulhu region, frost is observed almost exclusively on north-facing slopes, with a median slope of 12.8\textdegree~$\pm$~8\textdegree~at~1$\sigma$. The median projected slope at which frost is detected is 10\textdegree~$\pm$~8\textdegree~at~1$\sigma$. The higher slope angles associated with frost in Cthulhu suggest that these deposits are highly unstable and that their persistence may be governed by localized microclimate effects, which we explore in the following section.

\subsection{Simulation of the frost formation and evolution on the slopes}
\subsubsection{Baseline results}
\begin{figure*}[h!] \centering \includegraphics[width=0.95\textwidth]{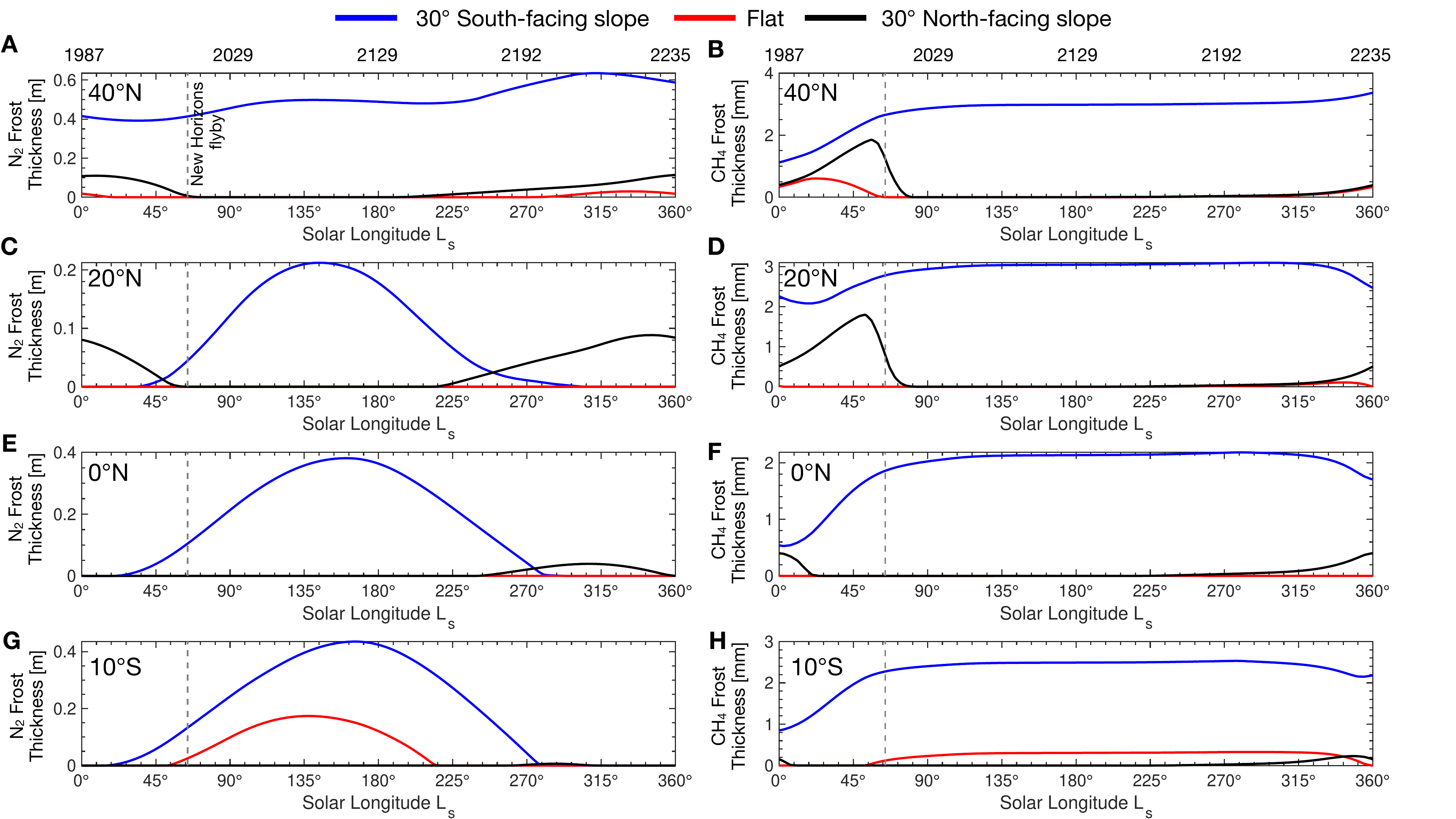} \caption{Seasonal evolution of frost on 30\textdegree~south-facing slopes (blue), flat surfaces (red), and 30\textdegree~north-facing slopes (black) on Pluto, shown for latitudes 40\textdegree N (A, B), 20\textdegree N (C, D), 0\textdegree N (E, F), and 10\textdegree S (G, H). For each latitude, N$_2$ frost thickness is shown in the left panels, and CH$_4$ frost thickness in the right panels. The dashed gray line marks the date of the New Horizons flyby. For cases with perennial accumulations, thickness values have been rescaled for clarity. The frost thicknesses shown here are zonal averages of the model outputs between 40\textdegree E and 150\textdegree E, corresponding to the Cthulhu region. } 
    \label{fig:allfrost_baseline}
    \end{figure*}

The seasonal evolution of CH$_4$ and N$_2$ frosts on flat surfaces, and on south and north-facing slopes, using the baseline model parameters described in Table \ref{table:inputs}, is shown in Fig. \ref{fig:allfrost_baseline}. These results highlight the significant influence of slope orientation on the condensation and sublimation of N$_2$ and CH$_4$. At high latitudes (Figs. \ref{fig:allfrost_baseline}a and \ref{fig:allfrost_baseline}b), the amount of N$_2$ frost forming on 30\textdegree~north-facing slopes is about five times higher (three times for CH$_4$) than on flat terrain. Furthermore, our model predicts perennial accumulation on 30\textdegree~south-facing slopes, whereas only seasonal frost is predicted on flat surfaces. At 20\textdegree N and at the equator (Figs. \ref{fig:allfrost_baseline}c–\ref{fig:allfrost_baseline}h), N$_2$ and CH$_4$ frosts form almost exclusively on slopes, with only negligible amounts on flat terrain. In the southern hemisphere (10\textdegree S; Figs. \ref{fig:allfrost_baseline}g and \ref{fig:allfrost_baseline}h), frost is again predicted on south-facing slopes and flat surfaces during southern winter, but not during northern winter, while our model also predicts frost formation on north-facing slopes.

These patterns can be explained by the differential solar illumination received by surfaces with different orientations. As shown by \citet[see their supplementary Figs. 8 and 9]{Bertrand2020}, 30\textdegree~north-facing slopes in the northern hemisphere receive about 35\% less insolation than flat terrain during winter and spring (and conversely for 30\textdegree~south-facing slopes). Volatiles should therefore accumulate preferentially on these cold traps relative to warmer surfaces. The seasonal evolution of frost deposits also follows the illumination cycle: frost forms in late autumn and winter, when slopes are in shadow, and sublimates in spring and summer, when they are exposed to sunlight. An important asymmetry is observed, however: Frost never completely disappears from south-facing slopes, unlike north-facing slopes (although the latter is highly dependent on the assumed albedo; see Sect. \ref{ssec:modelsensitivity}). In contrast, \cite{Bertrand2020} suggested, based on illumination computations, that frost on south-facing slopes should disappear after southern spring, when these slopes warm. The main difference between our model and theirs is the inclusion of the seasonal CH$_4$ cycle, which strongly affects the sublimation rate of CH$_4$ frost. As shown by \citet[see their Fig. 2]{Bertrand2016}, the maximum atmospheric CH$_4$ mixing ratio during southern summer is five times higher than during northern summer, because the atmosphere is then ten times thinner and provides less dilution of methane. As expressed in Eq. \ref{eq:dmdtCH$_4$}, the sublimation rate of CH$_4$ frost is highly sensitive to the CH$_4$ mixing ratio. This asymmetry results in a sublimation flux up to 3-6 times lower during the southern summer than in the northern summer, thereby favoring the formation of perennial CH$_4$ frost deposits on south-facing slopes. There is also a seasonal asymmetry in the atmospheric -and methane- density $\rho$, with larger values during the northern spring compared to the southern spring. This induces stronger sublimation fluxes in the northern spring on warm, north-facing slope, and, conversely, enhanced CH$_4$ condensation and accumulation on cold, south-facing slopes at the same periods, again favoring the preservation of frost and the formation of a perennial deposit. Once such a perennial deposit forms, its high albedo reduces the absorbed energy and the annual mean surface temperature, which in turn promotes further frost accumulation, establishing a positive feedback. Finally, although the Pluto orbit is highly eccentric, surface temperatures show no discernible asymmetry, suggesting that present-day CH$_4$ frost dynamics are governed mainly by atmospheric variations of pressure, density, and methane mixing ratio rather than by direct thermal effects. Indeed, as shown by \citet[see their Fig. S8]{Bertrand2020}, north and south-facing slopes receive nearly the same amount of solar insolation during spring and summer, when illuminated, and thus reach comparable temperatures. The reason is that the equinox and perihelion on Pluto occur close in time, which produces polar nights of equal length and summers of similar duration and intensity in both hemispheres \citep{Earle2017}. During other epochs, the timing of equinox relative to perihelion differs \citep{Dobrovolskis1997}. As a result, the high orbital eccentricity of Pluto produces substantial variations in seasonal illumination \citep{Earle2017} which may more importantly impact  the formation and sublimation of N$_2$ and CH$_4$ deposits .and mitigate the atmospheric effects described here \citep{Bertrand2018,Bertrand2019}.

    \subsubsection{Model sensitivity \label{ssec:modelsensitivity}}
\begin{figure*}[h!] \centering \includegraphics[width=0.9\textwidth]{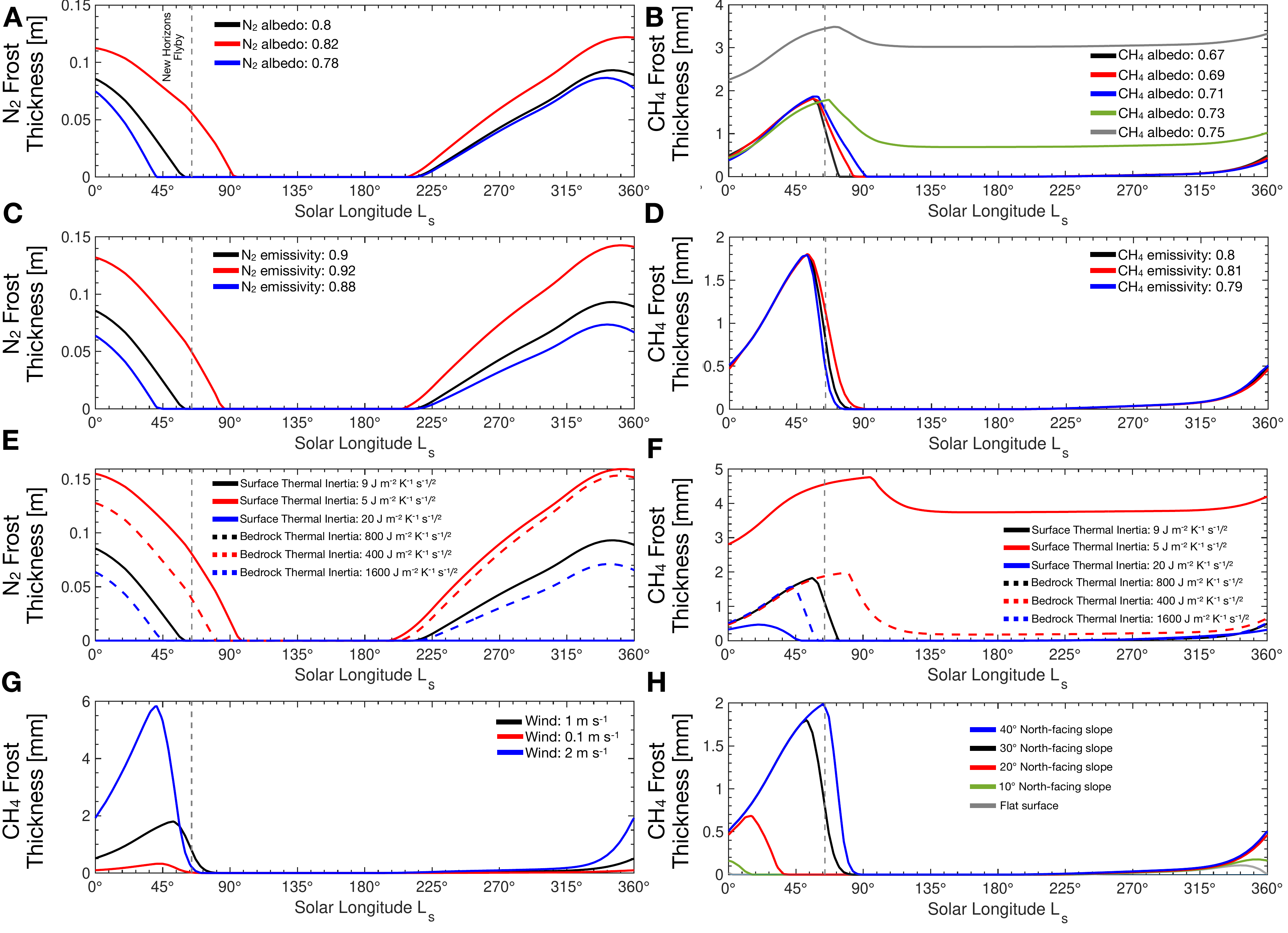} \caption{Sensitivity of the N$_2$ and CH$_4$ frost thicknesses to the model inputs: A) N$_2$ ice albedo; B) CH$_4$ ice albedo; C) N$_2$ ice emissivity; D) CH$_4$ ice emissivity;  E and F) thick lines: surface thermal inertia (first 5 cm), dashed lines: bedrock (below 5 cm) thermal inertia  G) wind speed H) slope angles.    All of the frost thicknesses are for a 30\textdegree~north-facing slope, located at 20\textdegree N. } 
    \label{fig:sensitivity}
    \end{figure*}

Figure \ref{fig:sensitivity} illustrates the sensitivity of our model results to the various parameters described in Table \ref{table:inputs}. As expected, increasing the ice albedo enhances the stability of frost (either N$_2$ or CH$_4$) during spring and summer (by about 20 Earth years for the case of N$_2$, Fig. \ref{fig:sensitivity}a, and $\sim$7 years for CH$_4$, Fig. \ref{fig:sensitivity}b). The stability of CH$_4$ frost is highly sensitive to its albedo \cite[c.f. the albedo feedback effect described in][]{Bertrand2020}: increasing the ice albedo from 0.71 to 0.73 can lead to the formation of perennial ice deposits. Increasing the emissivity of the ice also enhances frost stability with a similar magnitude (Figs. \ref{fig:sensitivity}c and \ref{fig:sensitivity}d), as it improves ice cooling and thus reduces the sublimation rate.  Frost deposits are also sensitive to the thermal conductivity of the surface and underlying bedrock (here defined at depths greater than 5 cm), and thereby to thermal inertia. High thermal inertia surfaces are more conductive and release the heat stored during daytime/summer into the cold seasons, which reduces the amount of ice that can condense and hence decreases its stability.   As highlighted by Eq. \ref{eq:dmdtCH$_4$} and Fig. \ref{fig:sensitivity}g, the condensation/sublimation rate of CH$_4$ ice is strongly controlled by wind speed: strong winds promote deposition but also accelerate sublimation, while weak winds inhibit both condensation and sublimation. In the latter case, however, given the thinness of the deposits, frost disappears faster than under high-wind conditions.  Finally, slope angle also affects frost stability: steeper slopes receive lower mean annual insolation, leading to lower surface temperatures and thus favoring the formation of thicker and more persistent frost deposits (Fig. \ref{fig:sensitivity}h). Additional sensitivity tests (not shown) indicate that bare surface albedo and emissivity affect the amount of heat absorbed by the ground during the day and its cooling at night, respectively. These processes influence the diurnal formation of frost and potentially its seasonal persistence. The amount of CH$_4$ in the atmosphere, and hence its transport, also affects the stability of frost deposits, as previously explained (see last paragraph). 

Our sensitivity experiments show that the timing of frost formation and disappearance is highly sensitive to the choice of key parameters, notably albedo, emissivity, diurnal thermal inertia, and regolith thermal inertia. These parameters have also been identified as critical for the evolution of volatile cycles across Pluto \citep[e.g., ][]{Bertrand2016,Bertrand2018,Bertrand2019,Young2021}. On Mars, the seasonality of CO$_2$ and H$_2$O frosts has been used to constrain the vertical stratification of the subsurface and its properties (notably the thermal inertia; \citealt{Vincendon2010,Vincendon2010water}).  Such an analysis is not possible here because of the strong multi-parameter sensitivity of our problem and the fact that we only have a single snapshot of frost dynamics, which leaves the system poorly constrained. Interestingly, Fig. \ref{fig:sensitivity} shows that the sublimation rate of CH$_4$ ice deposits is more sensitive to their albedo than to emissivity. In our simulations, at typical CH$_4$ frost temperatures, the infrared radiative cooling flux of the surface is two to three times weaker than the peak daytime insolation, explaining why albedo is the main driver of the sublimation rate.

\subsection{Comparison to New Horizons observations}

    \begin{figure*}[h!] \centering \includegraphics[width=0.9\textwidth]{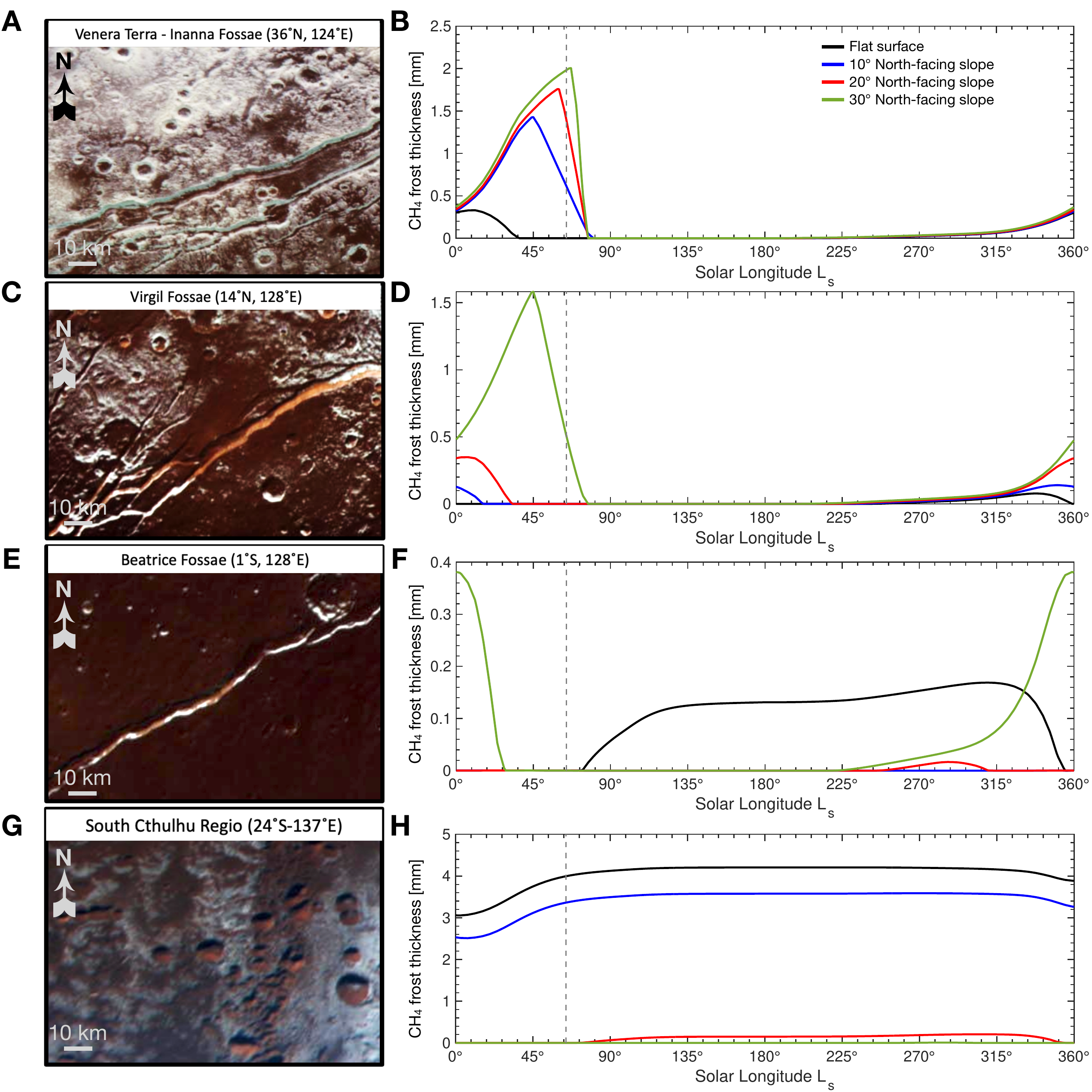} \caption{Comparisons between frost observations on Ralph/MVIC color images in the Cthulhu region reported by \cite{Bertrand2020}  at Venera Terra (A), Virgil Fossae (C), Beatrice Fossae (E), and the southern part of the region (G) and the model outputs (right) assuming an albedo of 0.69 for methane ice to achieve better agreement with the observations. For the performance of the baseline model (albedo of 0.67); see Fig. \ref{fig:compbertrand_067}. Perennial accumulation occurs on south-facing slopes (not shown). Perennial accumulation occurs on south-facing slopes (not shown). Our simulations predict that the CH$_4$ frost deposits (white patches  in panels A, C, E, G) observed on north-facing slopes correspond to seasonal CH$_4$ frost that gradually sublimates and should disappear shortly after the New Horizons flyby. No curves are shown for south-facing slopes, as no CH$_4$ frost is visible in panels A, C, E, G due to shadowing. These slopes also display only perennial accumulation, without seasonal variability (see Fig. \ref{fig:allfrost_baseline}). } 
    \label{fig:compbertrand_069}
    \end{figure*}

The seasonal evolution of methane frost predicted by our model is consistent with observations reported by \cite{Earle2022} and \citet[see our Figs. \ref{fig:allfrost_baseline} and \ref{fig:compbertrand_069}]{Bertrand2020}. At high latitudes (above 40\textdegree), CH$_4$ is predicted to be present on both north and south-facing slopes, consistent with the detections of \citet[see our Figs. \ref{fig:allfrost_baseline}a, \ref{fig:allfrost_baseline}b, and \ref{fig:compbertrand_069}b]{Earle2022}. Thick N$_2$ frost is also expected (Fig. \ref{fig:allfrost_baseline}a), in agreement with the observed mixtures of N$_2$ and CH$_4$ on these slopes.

At lower latitudes (below 20\textdegree), CH$_4$ frost is predicted to be always present on south-facing slopes, possibly diluted with N$_2$ ice that begins to condense (Figs. \ref{fig:allfrost_baseline}b to \ref{fig:allfrost_baseline}h and Figs. \ref{fig:compbertrand_069}b, \ref{fig:compbertrand_069}d, \ref{fig:compbertrand_069}f,  and \ref{fig:compbertrand_069}g). Neither \cite{Earle2022} nor \cite{Bertrand2020} detected frost on these south-facing slopes, however, although the very low illumination of these surfaces (below 0.3~W~m$^{-2}$; see Fig.~S8c of \citealt{Bertrand2020}) might have prevented detection. At 15\textdegree N, methane frost is predicted to persist on the steepest north-facing slopes (at least 30\textdegree~according to our model; Figs. \ref{fig:allfrost_baseline}d, \ref{fig:allfrost_baseline}f, \ref{fig:compbertrand_069}d, and \ref{fig:compbertrand_069}f) and to be sublimating. This is consistent with the reddish patterns observed next to ice deposits on such slopes, which were interpreted by \cite{Bertrand2020} as bedrock exposures indicating ongoing  frost sublimation.

The exact timing of frost sublimation remains highly sensitive to surface properties and local conditions (e.g., lower thermal inertia, higher slope angle), which can promote frost stability and explain the observed variability in frost persistence (Fig.~\ref{fig:sensitivity}). Notably, it appears that using a CH$_4$ ice albedo of 0.69 leads to a better agreement than using the baseline value of 0.67 to explain the persistence of these frosts (Figs. \ref{fig:compbertrand_069} and \ref{fig:compbertrand_067} ). In Sect.~\ref{ssec:resultfrostobs} we reported that frost was typically found on slopes of 12.8\textdegree~$\pm$~8\textdegree~at~1$\sigma$, but this value may be underestimated due to the limited resolution of the topographic dataset, which likely smooths slope gradients and reduces the apparent slope angles where frost is found.

At the equator, CH$_4$ frost has been observed on parts of north-facing slopes at Beatrice Fossae (Fig.~\ref{fig:compbertrand_069}e), while our model predicts that these deposits should have disappeared by the 2000s (Fig.~\ref{fig:compbertrand_069}f). The observed frosts are adjacent to reddish, frost-free terrains, again suggesting that local conditions strongly control frost persistence. \citet[see their Fig. 23]{Schenk2018} showed that the fracture wall scarps have typical slope angle between 35\textdegree~and 45\textdegree, and can locally reach up to  55\textdegree.  Assuming a slope angle of 40\textdegree, our model simulates perennial accumulation of CH$_4$ frost at this latitude (not shown).

Finally, the model predicts the formation of both N$_2$ and CH$_4$ frosts in the southern hemisphere as this region enters winter. Frost is found only on flat surfaces and south-facing slopes, consistent with the observations of \cite{Bertrand2020}.

\section{Discussions}
\subsection{Effect of the slope microclimates on the volatile cycles of Pluto \label{sec:discussion}}

Using a similar subgrid-slope parameterization in the Mars Planetary Climate Model, \cite{Lange2023} showed that slope microclimates exert little influence on the global seasonal cycles of CO$_2$ and H$_2$O. This is primarily because sloped terrains (here defined as surfaces with a slope angle greater than 5\textdegree) account for only about 5\% of Mars' total area, so frost and ice on flat regions dominate the volatile cycles. In contrast, nearly 30\% of the observed Pluto surface is non-flat (Fig. \ref{fig:histmu}), raising the question of how N$_2$ and CH$_4$ condensation on slopes might affect the global climate on Pluto. To quantify this effect, we run two simulations: one where the atmosphere interacts only with flat subgrid surfaces (all slope cover-fractions $\delta_i$ are set to 0 in Eqs. \ref{eq:traceurs} and \ref{eq:Tsurf}, called the smooth simulation), and one where $\delta_i$ values in every grid cell are set to those presented in Fig. \ref{fig:histmu} (the 'rough simulation'). The comparison between these simulations shows that slope microclimates do not significantly affect the N$_2$ cycle. The rough simulation yields a total pressure of 1.14~Pa in 2015, versus 1.2~Pa for the smooth simulation. As shown in Fig. \ref{fig:allfrost_baseline}, N$_2$ frost begins to form on south-facing slopes rather than on flat surfaces, affecting the amount of N$_2$ present in the atmosphere. Thick N$_2$ frost, notably thicker than on flat surfaces, can form by the time of New Horizons flyby (Fig. \ref{fig:allfrost_baseline}c), sometimes perennially, thereby reducing atmospheric N$_2$ and total pressure. During southern summer (Ls~=~270\textdegree), the total pressure is higher in the rough simulation than in the smooth simulation. As highlighted in Fig. \ref{fig:allfrost_baseline}, N$_2$ frost on south-facing slopes sublimes, replenishing the atmosphere and explaining the increased surface pressure compared to the smooth case where N$_2$ frost begins forming at the surface.  This comparison indicates that, although the Pluto surface is more strongly sloped than Mars and N$_2$–CH$_4$ frost can develop on inclined terrains where it cannot on flats, slope microclimates still appear to exert only a limited influence on the N$_2$ seasonal cycle on Pluto. Note that  in our simulations, we adopted a uniform slope distribution within each model grid cell to avoid the computational constraints discussed in Sect. \ref{ssec:slopeinVTM}. Actually, certain regions, especially those covered by thick, perennial N$_2$ ice, are likely smoother than this idealization. In addition, the Pluto VTM treats atmospheric circulation and volatile transport in a simplified manner. Hence, a robust quantitative evaluation of slope‐driven microclimates on the seasonal cycles on Pluto will  require a complete topographic map combined with simulations using the Pluto Global Climate Model, which we defer to future work.

The methane cycle diagnostic is more complex. The perennial accumulation of CH$_4$ frost on south-facing slopes and thicker frost on north-facing slopes compared to flat surfaces suggest that the rough simulation should exhibit lower atmospheric CH$_4$ volume-mixing ratios than the smooth case. The different total pressures between the simulations bias the results, however. Due to lower pressure in the rough simulation during northern summer, methane is less diluted, resulting in higher atmospheric CH$_4$ volume-mixing ratios. Conversely, higher pressure during the southern summer leads to greater methane dilution and lower mixing ratios compared to the smooth simulation. Since pressure variations also impact the methane cycle and the amount of methane condensing and subliming, this bias is not straightforward to correct and prevents us from drawing definitive conclusions about slope microclimate effects on the methane cycle.

\subsection{Evolution of the Cthulhu region in the coming decades}
\label{ssec:futureobs}
As illustrated in Figs. \ref{fig:allfrost_baseline} and \ref{fig:compbertrand_069},  CH$_4$ and N$_2$ frosts are expected to disappear completely from slopes and flat terrains between 40\textdegree~and 0\textdegree~N in the Cthulhu region by around 2030. N$_2$ frosts should instead begin to accumulate on the permanent CH$_4$ ice deposits located on south-facing slopes. Conversely, little to no N$_2$  or CH$_4$ frost (thickness $<$~1~mm; Fig. \ref{fig:allfrost_baseline} ) is expected to form on the flat terrains of Cthulhu. As a result, this region should retain its dark albedo as mapped by \cite{Hofgartner2023} over the coming decades to centuries.

In the southern hemisphere, N$_2$  frosts a few tens of centimeters thick and a thin layer of methane frost ($<$~0.1~mm) are predicted to form by 2100. Although our estimates of frost amounts remain sensitive to the model parameters and simplifications (see Sects. \ref{ssec:slopeinVTM}, \ref{ssec:limits}, and \ref{ssec:modelsensitivity}), we can reasonably anticipate that the northern mid-latitudes will continue to sublimate and appear darker, whereas the southern hemisphere will gradually brighten.

\section{Conclusions \label{sec:conclusions}}

We analyzed the frost on the slopes in the Cthulhu region of Pluto, which was previously thought to be frost-free because its surface is dark, and we developed a parameterization in the Pluto Volatile Transport Model to explain this and study the stability of the frost. The main conclusions of our investigation are listed below.

\begin{itemize}
    \item CH$_4$ frost is widespread on north-facing slopes of the Cthulhu region (Fig. \ref{fig:mapfrost}), generally on steep slopes (mean slope angle 12.8\textdegree~$\pm$~8\textdegree~at~1$\sigma$; Fig. \ref{fig:frostdetectedstat}). No frost is observed on south-facing slopes, but this may be a consequence of their very low insolation (<0.3~W~m$^{-2}$) during the period of the Pluto orbit we considered, which might prevent detection.

    \item The volatile transport model for Pluto, coupled with the subgrid-slope parameterization (Fig. \ref{fig:illustrationparam}), showed that frost on north-facing slopes corresponds to seasonal frost that currently sublimates at the time of the New Horizons observations. This agrees well with the data (Figs. \ref{fig:allfrost_baseline} and \ref{fig:compbertrand_069}). The sublimation rates and frost persistence throughout the Pluto orbit are highly sensitive to surface properties such as the ice albedo, the surface, the bedrock thermal inertia, and the slope angle (Fig. \ref{fig:sensitivity}).

\item N$_2$ frost deposits can form on north- and south-facing slopes during winter in the Cthulhu region, whereas they are not expected to develop on flat surfaces (Fig. \ref{fig:allfrost_baseline}). At the time of the New Horizons flyby, N$_2$ frost on north-facing slopes should already have sublimated, while accumulation is expected to begin on south-facing slopes and reach thicknesses of a few centimeters (Fig. \ref{fig:allfrost_baseline}).

    \item Perennial CH$_4$ frost deposits are expected to form on south-facing slopes (Figs. \ref{fig:allfrost_baseline} and \ref{fig:compbertrand_069}), as sublimation during southern summer is mitigated by the higher CH$_4$ volume-mixing ratio that prevails during this season.

    \item Despite the abundant slopes on Pluto and the ability of frost to form (either seasonally or perennially) on inclined terrains where it cannot accumulate on flat surfaces, the slope microclimates appear to play only a minor role in the global volatile cycles on Pluto (Sect. \ref{sec:discussion}). Conversely, the global volatile cycle strongly shapes local slope microclimates by controlling the amount of condensable material available for the formation of frost. A confirmation of the interactions between local slope-microclimates and the large-scale volatile cycles will require a complete climate model and a detailed map of the global topography of Pluto.

    \item By about 2030, the CH$_4$ and N$_2$ frosts are expected to disappear from slopes and flat terrains between 0\textdegree~and 40\textdegree N in the Cthulhu region, while N$_2$ frost will start to accumulate on south-facing perennial deposits of CH$_4$ ice. Hence, the Cthulhu region is expected to conserve its dark appearance during the next centuries.  In the southern hemisphere, N$_2$ frosts up to a few dozen centimeters and a thin CH$_4$ layer are predicted by 2100, suggesting that the northern mid-latitudes will remain dark, while the south brightens gradually (Sect. \ref{ssec:futureobs}).

    \item Our model generally predicts the seasonal formation and disappearance of CH$_4$ frost within the Cthulhu region. In particular, it reproduces the frost on north-facing slopes down to about 10\textdegree N at the time of the New Horizons flyby. These discrepancies likely stem from the model constraint, which limits slope angles to 30\textdegree, whereas local slopes can reach up to 40\textdegree~along fault scarps and crater walls. When the maximum slope angle was increased in sensitivity tests, the modeled frost distribution became consistent with the observations. Future high-resolution observations might confirm perennial frost on south-facing slopes. On the other hand, new James Webb Space Telescope observations of the northern and southern Pluto hemispheres could test our prediction that the northern frosts are progressively vanishing as the southern hemisphere enters winter, and this might help us to constrain key model parameters, such as the optical properties of the ice and the thermal inertia of the substrate \citep[e.g.,][]{Bertrand2025}, which are currently poorly constrained.
\end{itemize}

While the subgrid-slope model we developed was only applied to Mars and Pluto so far, the generality of our parameterization makes it suitable for studying volatile condensation and sublimation on other airless bodies or planets with thin atmospheres. In particular, coupling this parameterization with the volatile transport model for Triton \citep{Bertrand2022} would allow us to investigate the formation of ice deposits on slopes, and potentially, the effect of slopes on seasonal volatile cycles. This would allow us to generalize the conclusions drawn here.

\begin{acknowledgements}
      LL's research was supported by an appointment to the NASA Postdoctoral Program  administered by Oak Ridge Associated Universities under contract with NASA at the Jet Propulsion Laboratory,  California Institute of Technology, under a contract with the National Aeronautics and Space Administration (80NM0018D0004). © 2025. All rights reserved.  L.A. Young's was funded by NASA grant NFDAP 80NSSC23K0666. Other authors acknowledge the support of the French Agence Nationale de la Recherche (ANR), under grant ANR-23-CE49-0006 (project SHERPAS). 

      The model used in this study can be retrieved at: \url{https://trac.lmd.jussieu.fr/Planeto/browser/trunk/LMDZ.PLUTO}.   The topographic Pluto data we used are taken from \cite{Schenk2018}.  CH$_4$ band depth map and phase index retrieved from the LEISA instrument are from \cite{Schmitt2017} and \cite{Gabasova2021}. Albedo maps are taken from \citep{Hofgartner2023}. Data files used to generate the figures in this analysis are available in a public repository, see \cite{Lange2025plutodata}. 
\end{acknowledgements}

\bibliographystyle{aa} 
\bibliography{plutobib} 

@article{Stern2015,
  title = "{The Pluto system: Initial results from its exploration by New Horizons}",
  volume = {350},
  ISSN = {1095-9203},
  DOI = {10.1126/science.aad1815},
  number = {6258},
  journal = {Science},
  publisher = {American Association for the Advancement of Science (AAAS)},
  author = {Stern,  S. A. and Bagenal,  F. and Ennico,  K. and Gladstone,  G. R. and Grundy,  W. M. and McKinnon,  W. B. and Moore,  J. M. and Olkin,  C. B. and Spencer,  J. R. and Weaver,  H. A. and Young,  L. A. and Andert,  T. and Andrews,  J. and Banks,  M. and Bauer,  B. and Bauman,  J. and Barnouin,  O. S. and Bedini,  P. and Beisser,  K. and Beyer,  R. A. and Bhaskaran,  S. and Binzel,  R. P. and Birath,  E. and Bird,  M. and Bogan,  D. J. and Bowman,  A. and Bray,  V. J. and Brozovic,  M. and Bryan,  C. and Buckley,  M. R. and Buie,  M. W. and Buratti,  B. J. and Bushman,  S. S. and Calloway,  A. and Carcich,  B. and Cheng,  A. F. and Conard,  S. and Conrad,  C. A. and Cook,  J. C. and Cruikshank,  D. P. and Custodio,  O. S. and Dalle Ore,  C. M. and Deboy,  C. and Dischner,  Z. J. B. and Dumont,  P. and Earle,  A. M. and Elliott,  H. A. and Ercol,  J. and Ernst,  C. M. and Finley,  T. and Flanigan,  S. H. and Fountain,  G. and Freeze,  M. J. and Greathouse,  T. and Green,  J. L. and Guo,  Y. and Hahn,  M. and Hamilton,  D. P. and Hamilton,  S. A. and Hanley,  J. and Harch,  A. and Hart,  H. M. and Hersman,  C. B. and Hill,  A. and Hill,  M. E. and Hinson,  D. P. and Holdridge,  M. E. and Horanyi,  M. and Howard,  A. D. and Howett,  C. J. A. and Jackman,  C. and Jacobson,  R. A. and Jennings,  D. E. and Kammer,  J. A. and Kang,  H. K. and Kaufmann,  D. E. and Kollmann,  P. and Krimigis,  S. M. and Kusnierkiewicz,  D. and Lauer,  T. R. and Lee,  J. E. and Lindstrom,  K. L. and Linscott,  I. R. and Lisse,  C. M. and Lunsford,  A. W. and Mallder,  V. A. and Martin,  N. and McComas,  D. J. and McNutt,  R. L. and Mehoke,  D. and Mehoke,  T. and Melin,  E. D. and Mutchler,  M. and Nelson,  D. and Nimmo,  F. and Nunez,  J. I. and Ocampo,  A. and Owen,  W. M. and Paetzold,  M. and Page,  B. and Parker,  A. H. and Parker,  J. W. and Pelletier,  F. and Peterson,  J. and Pinkine,  N. and Piquette,  M. and Porter,  S. B. and Protopapa,  S. and Redfern,  J. and Reitsema,  H. J. and Reuter,  D. C. and Roberts,  J. H. and Robbins,  S. J. and Rogers,  G. and Rose,  D. and Runyon,  K. and Retherford,  K. D. and Ryschkewitsch,  M. G. and Schenk,  P. and Schindhelm,  E. and Sepan,  B. and Showalter,  M. R. and Singer,  K. N. and Soluri,  M. and Stanbridge,  D. and Steffl,  A. J. and Strobel,  D. F. and Stryk,  T. and Summers,  M. E. and Szalay,  J. R. and Tapley,  M. and Taylor,  A. and Taylor,  H. and Throop,  H. B. and Tsang,  C. C. C. and Tyler,  G. L. and Umurhan,  O. M. and Verbiscer,  A. J. and Versteeg,  M. H. and Vincent,  M. and Webbert,  R. and Weidner,  S. and Weigle,  G. E. and White,  O. L. and Whittenburg,  K. and Williams,  B. G. and Williams,  K. and Williams,  S. and Woods,  W. W. and Zangari,  A. M. and Zirnstein,  E.},
  year = {2015},
}

@article{Grundy2016,
  title = "{Surface compositions across Pluto and Charon}",
  volume = {351},
  ISSN = {1095-9203},
  DOI = {10.1126/science.aad9189},
  number = {6279},
  journal = {Science},
  publisher = {American Association for the Advancement of Science (AAAS)},
  author = {Grundy,  W. M. and Binzel,  R. P. and Buratti,  B. J. and Cook,  J. C. and Cruikshank,  D. P. and Dalle Ore,  C. M. and Earle,  A. M. and Ennico,  K. and Howett,  C. J. A. and Lunsford,  A. W. and Olkin,  C. B. and Parker,  A. H. and Philippe,  S. and Protopapa,  S. and Quirico,  E. and Reuter,  D. C. and Schmitt,  B. and Singer,  K. N. and Verbiscer,  A. J. and Beyer,  R. A. and Buie,  M. W. and Cheng,  A. F. and Jennings,  D. E. and Linscott,  I. R. and Parker,  J. Wm. and Schenk,  P. M. and Spencer,  J. R. and Stansberry,  J. A. and Stern,  S. A. and Throop,  H. B. and Tsang,  C. C. C. and Weaver,  H. A. and Weigle,  G. E. and Young,  L. A.},
  year = {2016},
}

@article{Moore2016,
  title = "{The geology of Pluto and Charon through the eyes of New Horizons}",
  volume = {351},
  ISSN = {1095-9203},
  DOI = {10.1126/science.aad7055},
  number = {6279},
  journal = {Science},
  publisher = {American Association for the Advancement of Science (AAAS)},
  author = {Moore,  Jeffrey M. and McKinnon,  William B. and Spencer,  John R. and Howard,  Alan D. and Schenk,  Paul M. and Beyer,  Ross A. and Nimmo,  Francis and Singer,  Kelsi N. and Umurhan,  Orkan M. and White,  Oliver L. and Stern,  S. Alan and Ennico,  Kimberly and Olkin,  Cathy B. and Weaver,  Harold A. and Young,  Leslie A. and Binzel,  Richard P. and Buie,  Marc W. and Buratti,  Bonnie J. and Cheng,  Andrew F. and Cruikshank,  Dale P. and Grundy,  Will M. and Linscott,  Ivan R. and Reitsema,  Harold J. and Reuter,  Dennis C. and Showalter,  Mark R. and Bray,  Veronica J. and Chavez,  Carrie L. and Howett,  Carly J. A. and Lauer,  Tod R. and Lisse,  Carey M. and Parker,  Alex Harrison and Porter,  S. B. and Robbins,  Stuart J. and Runyon,  Kirby and Stryk,  Ted and Throop,  Henry B. and Tsang,  Constantine C. C. and Verbiscer,  Anne J. and Zangari,  Amanda M. and Chaikin,  Andrew L. and Wilhelms,  Don E. and Bagenal,  F. and Gladstone,  G. R. and Andert,  T. and Andrews,  J. and Banks,  M. and Bauer,  B. and Bauman,  J. and Barnouin,  O. S. and Bedini,  P. and Beisser,  K. and Bhaskaran,  S. and Birath,  E. and Bird,  M. and Bogan,  D. J. and Bowman,  A. and Brozovic,  M. and Bryan,  C. and Buckley,  M. R. and Bushman,  S. S. and Calloway,  A. and Carcich,  B. and Conard,  S. and Conrad,  C. A. and Cook,  J. C. and Custodio,  O. S. and Ore,  C. M. Dalle and Deboy,  C. and Dischner,  Z. J. B. and Dumont,  P. and Earle,  A. M. and Elliott,  H. A. and Ercol,  J. and Ernst,  C. M. and Finley,  T. and Flanigan,  S. H. and Fountain,  G. and Freeze,  M. J. and Greathouse,  T. and Green,  J. L. and Guo,  Y. and Hahn,  M. and Hamilton,  D. P. and Hamilton,  S. A. and Hanley,  J. and Harch,  A. and Hart,  H. M. and Hersman,  C. B. and Hill,  A. and Hill,  M. E. and Hinson,  D. P. and Holdridge,  M. E. and Horanyi,  M. and Jackman,  C. and Jacobson,  R. A. and Jennings,  D. E. and Kammer,  J. A. and Kang,  H. K. and Kaufmann,  D. E. and Kollmann,  P. and Krimigis,  S. M. and Kusnierkiewicz,  D. and Lee,  J. E. and Lindstrom,  K. L. and Lunsford,  A. W. and Mallder,  V. A. and Martin,  N. and McComas,  D. J. and McNutt,  R. L. and Mehoke,  D. and Mehoke,  T. and Melin,  E. D. and Mutchler,  M. and Nelson,  D. and Nunez,  J. I. and Ocampo,  A. and Owen,  W. M. and Paetzold,  M. and Page,  B. and Parker,  J. W. and Pelletier,  F. and Peterson,  J. and Pinkine,  N. and Piquette,  M. and Protopapa,  S. and Redfern,  J. and Roberts,  J. H. and Rogers,  G. and Rose,  D. and Retherford,  K. D. and Ryschkewitsch,  M. G. and Schindhelm,  E. and Sepan,  B. and Soluri,  M. and Stanbridge,  D. and Steffl,  A. J. and Strobel,  D. F. and Summers,  M. E. and Szalay,  J. R. and Tapley,  M. and Taylor,  A. and Taylor,  H. and Tyler,  G. L. and Versteeg,  M. H. and Vincent,  M. and Webbert,  R. and Weidner,  S. and Weigle,  G. E. and Whittenburg,  K. and Williams,  B. G. and Williams,  K. and Williams,  S. and Woods,  W. W. and Zirnstein,  E.},
  year = {2016},
  pages = {1284–1293}
}

@article{Schmitt2017,
  title = "{Physical state and distribution of materials at the surface of Pluto from New Horizons LEISA imaging spectrometer}",
  volume = {287},
  ISSN = {0019-1035},
  DOI = {10.1016/j.icarus.2016.12.025},
  journal = {Icarus},
  publisher = {Elsevier BV},
  author = {Schmitt,  B. and Philippe,  S. and Grundy,  W.M. and Reuter,  D.C. and C\^ote,  R. and Quirico,  E. and Protopapa,  S. and Young,  L.A. and Binzel,  R.P. and Cook,  J.C. and Cruikshank,  D.P. and Dalle Ore,  C.M. and Earle,  A.M. and Ennico,  K. and Howett,  C.J.A. and Jennings,  D.E. and Linscott,  I.R. and Lunsford,  A.W. and Olkin,  C.B. and Parker,  A.H. and Parker,  J.Wm. and Singer,  K.N. and Spencer,  J.R. and Stansberry,  J.A. and Stern,  S.A. and Tsang,  C.C.C. and Verbiscer,  A.J. and Weaver,  H.A.},
  year = {2017},
  pages = {229–260}
}

@article{Earle2018,
  title = "{Methane distribution on Pluto as mapped by the New Horizons Ralph/MVIC instrument}",
  volume = {314},
  ISSN = {0019-1035},
  DOI = {10.1016/j.icarus.2018.06.005},
  journal = {Icarus},
  publisher = {Elsevier BV},
  author = {Earle,  Alissa M. and Grundy,  W. and Howett,  C.J.A. and Olkin,  C.B. and Parker,  A.H. and Scipioni,  F. and Binzel,  R.P. and Beyer,  R.A. and Cook,  J.C. and Cruikshank,  D.P. and Dalle Ore,  C.M. and Ennico,  K. and Protopapa,  S. and Reuter,  D.C. and Schenk,  P.M. and Schmitt,  B. and Stern,  S.A. and Weaver,  H.A. and Young,  L.A.},
  year = {2018},
  pages = {195–209}
}

@article{Protopapa2017,
  title = "{Pluto’s global surface composition through pixel-by-pixel Hapke modeling of New Horizons Ralph/LEISA data}",
  volume = {287},
  ISSN = {0019-1035},
  DOI = {10.1016/j.icarus.2016.11.028},
  journal = {Icarus},
  publisher = {Elsevier BV},
  author = {Protopapa,  S. and Grundy,  W.M. and Reuter,  D.C. and Hamilton,  D.P. and Dalle Ore,  C.M. and Cook,  J.C. and Cruikshank,  D.P. and Schmitt,  B. and Philippe,  S. and Quirico,  E. and Binzel,  R.P. and Earle,  A.M. and Ennico,  K. and Howett,  C.J.A. and Lunsford,  A.W. and Olkin,  C.B. and Parker,  A. and Singer,  K.N. and Stern,  A. and Verbiscer,  A.J. and Weaver,  H.A. and Young,  L.A.},
  year = {2017},
  pages = {218–228}
}

@article{Buratti2017,
  title = "{Global albedos of Pluto and Charon from LORRI New Horizons observations}",
  volume = {287},
  ISSN = {0019-1035},
  DOI = {10.1016/j.icarus.2016.11.012},
  journal = {Icarus},
  publisher = {Elsevier BV},
  author = {Buratti,  B.J. and Hofgartner,  J.D. and Hicks,  M.D. and Weaver,  H.A. and Stern,  S.A. and Momary,  T. and Mosher,  J.A. and Beyer,  R.A. and Verbiscer,  A.J. and Zangari,  A.M. and Young,  L.A. and Lisse,  C.M. and Singer,  K. and Cheng,  A. and Grundy,  W. and Ennico,  K. and Olkin,  C.B.},
  year = {2017},
  pages = {207–217}
}

@article{Moore2018,
  title = "{Bladed Terrain on Pluto: Possible origins and evolution}",
  volume = {300},
  ISSN = {0019-1035},
  DOI = {10.1016/j.icarus.2017.08.031},
  journal = {Icarus},
  publisher = {Elsevier BV},
  author = {Moore,  Jeffrey M. and Howard,  Alan D. and Umurhan,  Orkan M. and White,  Oliver L. and Schenk,  Paul M. and Beyer,  Ross A. and McKinnon,  William B. and Spencer,  John R. and Singer,  Kelsi N. and Grundy,  William M. and Earle,  Alissa M. and Schmitt,  Bernard and Protopapa,  Silvia and Nimmo,  Francis and Cruikshank,  Dale P. and Hinson,  David P. and Young,  Leslie A. and Stern,  S. Alan and Weaver,  Harold A. and Olkin,  Cathy B. and Ennico,  Kimberly and Collins,  Geoffrey and Bertrand,  Tanguy and Forget,  Fran\c{c}ois and Scipioni,  Francesca},
  year = {2018},
  pages = {129–144}
}

@article{Hansen1996,
  title = "{Seasonal Nitrogen Cycles on Pluto}",
  volume = {120},
  ISSN = {0019-1035},
  DOI = {10.1006/icar.1996.0049},
  number = {2},
  journal = {Icarus},
  publisher = {Elsevier BV},
  author = {Hansen,  Candice J. and Paige,  David A.},
  year = {1996},
  pages = {247–265}
}

@article{Bertrand2018,
  title = "{The nitrogen cycles on Pluto over seasonal and astronomical timescales}",
  volume = {309},
  ISSN = {0019-1035},
  DOI = {10.1016/j.icarus.2018.03.012},
  journal = {Icarus},
  publisher = {Elsevier BV},
  author = {Bertrand,  T. and Forget,  F. and Umurhan,  O.M. and Grundy,  W.M. and Schmitt,  B. and Protopapa,  S. and Zangari,  A.M. and White,  O.L. and Schenk,  P.M. and Singer,  K.N. and Stern,  A. and Weaver,  H.A. and Young,  L.A. and Ennico,  K. and Olkin,  C.B.},
  year = {2018},
  pages = {277–296}
}

@article{Bertrand2019,
  title = "{The CH$_4$ cycles on Pluto over seasonal and astronomical timescales}",
  volume = {329},
  ISSN = {0019-1035},
  DOI = {10.1016/j.icarus.2019.02.007},
  journal = {Icarus},
  publisher = {Elsevier BV},
  author = {Bertrand,  T. and Forget,  F. and Umurhan,  O.M. and Moore,  J.M. and Young,  L.A. and Protopapa,  S. and Grundy,  W.M. and Schmitt,  B. and Dhingra,  R.D. and Binzel,  R.P. and Earle,  A.M. and Cruikshank,  D.P. and Stern,  S.A. and Weaver,  H.A. and Ennico,  K. and Olkin,  C.B.},
  year = {2019},
  pages = {148–165}
}

@article{Bertrand2020,
  title = "{Equatorial mountains on Pluto are covered by methane frosts resulting from a unique atmospheric process}",
  volume = {11},
  ISSN = {2041-1723},
  DOI = {10.1038/s41467-020-18845-3},
  number = {1},
  journal = {Nat. Comm.},
  publisher = {Springer Science and Business Media LLC},
  author = {Bertrand,  Tanguy and Forget,  Fran\c{c}ois and Schmitt,  Bernard and White,  Oliver L. and Grundy,  William M.},
  year = {2020},
}

@article{Young2012,
  title = "{Volatile transport on inhomogeneous surfaces: I – Analytic expressions,  with application to Pluto’s day}",
  volume = {221},
  ISSN = {0019-1035},
  DOI = {10.1016/j.icarus.2012.06.032},
  number = {1},
  journal = {Icarus},
  publisher = {Elsevier BV},
  author = {Young,  Leslie A.},
  year = {2012},
  pages = {80–88}
}

@article{Bertrand2016,
  title = "{Observed glacier and volatile distribution on Pluto from atmosphere–topography processes}",
  volume = {540},
  ISSN = {1476-4687},
  DOI = {10.1038/nature19337},
  number = {7631},
  journal = {Nature},
  publisher = {Springer Science and Business Media LLC},
  author = {Bertrand,  Tanguy and Forget,  Fran\c{c}ois},
  year = {2016},
  pages = {86–89}
}

@article{Young2013,
  title = "{PLUTO’S SEASONS: NEW PREDICTIONS FOR NEW HORIZONS}",
  volume = {766},
  ISSN = {2041-8213},
  DOI = {10.1088/2041-8205/766/2/l22},
  number = {2},
  journal = {ApJ},
  publisher = {American Astronomical Society},
  author = {Young,  L. A.},
  year = {2013},
  pages = {L22}
}

@article{Forget2017,
  title = "{A post-new horizons global climate model of Pluto including the N$_2$ ,  CH$_4$ and CO cycles}",
  volume = {287},
  ISSN = {0019-1035},
  DOI = {10.1016/j.icarus.2016.11.038},
  journal = {Icarus},
  publisher = {Elsevier BV},
  author = {Forget,  F. and Bertrand,  T. and Vangvichith,  M. and Leconte,  J. and Millour,  E. and Lellouch,  E.},
  year = {2017},
  pages = {54–71}
}

@article{Zalucha2013,
  title = "{A 3D general circulation model for Pluto and Triton with fixed volatile abundance and simplified surface forcing}",
  volume = {223},
  ISSN = {0019-1035},
  DOI = {10.1016/j.icarus.2013.01.026},
  number = {2},
  journal = {Icarus},
  publisher = {Elsevier BV},
  author = {Zalucha,  Angela M. and Michaels,  Timothy I.},
  year = {2013},
  pages = {819–831}
}

@article{Toigo2015,
  title = "{General circulation models of the dynamics of Pluto’s volatile transport on the eve of the New Horizons encounter}",
  volume = {254},
  ISSN = {0019-1035},
  DOI = {10.1016/j.icarus.2015.03.034},
  journal = {Icarus},
  publisher = {Elsevier BV},
  author = {Toigo,  Anthony D. and French,  Richard G. and Gierasch,  Peter J. and Guzewich,  Scott D. and Zhu,  Xun and Richardson,  Mark I.},
  year = {2015},
  pages = {306–323}
}

@inbook{Moore2021,
  title = "{The Landscapes of Pluto as Witness to Climate Evolution}",
  ISBN = {9780816540945},
  DOI = {10.2458/azu_uapress_9780816540945-ch006},
  booktitle = {The Pluto System After New Horizons},
  publisher = {University of Arizona Press},
  author = {Moore,  J. M.},
  year = {2021},
}

@INCOLLECTION{Young2021,
       author = {{Young}, L.~A. and {Bertrand}, T. and {Trafton}, L.~M. and {Forget}, F. and {Earle}, A.~M. and {Sicardy}, B.},
        title = "{Pluto's Volatile and Climate Cycles on Short and Long Timescales}",
    booktitle = {The Pluto System After New Horizons},
         year = 2021,
       editor = {{Stern}, S.~A. and {Moore}, J.~M. and {Grundy}, W.~M. and {Young}, L.~A. and {Binzel}, R.~P.},
        pages = {321-361},
          doi = {10.2458/azu_uapress_9780816540945-ch014},
     
}

@article{Earle2018alb,
  title = "{Albedo matters: Understanding runaway albedo variations on Pluto}",
  volume = {303},
  ISSN = {0019-1035},
  DOI = {10.1016/j.icarus.2017.12.015},
  journal = {Icarus},
  publisher = {Elsevier BV},
  author = {Earle,  Alissa M. and Binzel,  Richard P. and Young,  Leslie A. and Stern,  S.A. and Ennico,  K. and Grundy,  W. and Olkin,  C.B. and Weaver,  H.A.},
  year = {2018},
  pages = {1–9}
}

@article{Earle2022,
  title = "{Tracing seasonal trends across Pluto’s craters: New Horizons Ralph/MVIC results}",
  volume = {373},
  ISSN = {0019-1035},
  DOI = {10.1016/j.icarus.2021.114771},
  journal = {Icarus},
  publisher = {Elsevier BV},
  author = {Earle,  Alissa M. and Binzel,  R.P. and Keane,  J.T. and Grundy,  W.M. and Howett,  C.J.A. and Olkin,  C.B. and Parker,  A.H. and Scipioni,  F. and Ennico,  K. and Stern,  S.A. and Weaver,  H.A. and Young,  L.A.},
  year = {2022},
  pages = {114771}
}

@article{Lange2023,
  title = "{Modeling Slope Microclimates in the Mars Planetary Climate Model}",
  volume = {128},
  ISSN = {2169-9100},
  DOI = {10.1029/2023je007915},
  number = {10},
  journal = {J. Geophys. Res. (Planets)},
  publisher = {American Geophysical Union (AGU)},
  author = {Lange,  L. and Forget,  F. and Dupont,  E. and Vandemeulebrouck,  R. and Spiga,  A. and Millour,  E. and Vincendon,  M. and Bierjon,  A.},
  year = {2023},
}

@article{Lange2024,
  title = "{Observations of Water Frost on Mars With THEMIS: Application to the Presence of Brines and the Stability of (Sub)Surface Water Ice}",
  volume = {129},
  ISSN = {2169-9100},
  DOI = {10.1029/2024je008489},
  number = {10},
  journal = {J. Geophys. Res. (Planets)},
  publisher = {American Geophysical Union (AGU)},
  author = {Lange,  L. and Piqueux,  S. and Edwards,  C. S. and Forget,  F. and Naar,  J. and Vos,  E. and Szantai,  A.},
  year = {2024},
}

@article{Vincendon2010,
  doi = {10.1029/2009gl041426},
  year = {2010a},
  publisher = {American Geophysical Union ({AGU})},
  volume = {37},
  number = {1},
  author = {Mathieu {Vincendon} and John Mustard and Fran{\c{c}}ois Forget and Mikhail Kreslavsky and Aymeric Spiga and Scott Murchie and Jean-Pierre Bibring},
  title = {Near-tropical subsurface ice on {M}ars},
  journal = {GRL}
}

@ARTICLE{Vincendon2010water,
       author = {{Vincendon}, Mathieu and {Forget}, Fran{\c{c}}ois and {Mustard}, John},
        title = "{Water ice at low to midlatitudes on {M}ars}",
      journal = {J. Geophys. Res. (Planets)},
         year = {2010b},
       volume = {115},
       number = {E10},
          eid = {E10001},
        pages = {E10001},
          doi = {10.1029/2010JE003584},

}

@article{Bertrand2025,
  title = "{Evidence of haze control of Pluto’s atmospheric heat balance from JWST/MIRI thermal light curves}",
  ISSN = {2397-3366},
  DOI = {10.1038/s41550-025-02573-z},
  journal = {Nat. Astro.},
  publisher = {Springer Science and Business Media LLC},
  author = {Bertrand,  Tanguy and Lellouch,  Emmanuel and Holler,  Bryan and Stansberry,  John and Wong,  Ian and Zhang,  Xi and Lavvas,  Panayotis and Dufaux,  Elodie and Merlin,  Frederic and Villanueva,  Geronimo and Wan,  Linfeng and Pinilla-Alonso,  Noemí and de Souza Feliciano,  Ana Carolina and Murray,  Katherine},
  year = {2025},
}

@article{Lellouch2011,
  title = "{Thermal properties of Pluto’s and Charon’s surfaces from Spitzer observations}",
  volume = {214},
  ISSN = {0019-1035},
  DOI = {10.1016/j.icarus.2011.05.035},
  number = {2},
  journal = {Icarus},
  publisher = {Elsevier BV},
  author = {Lellouch,  Emmanuel and Stansberry,  John and Emery,  Josh and Grundy,  Will and Cruikshank,  Dale P.},
  year = {2011},
  pages = {701–716}
}

@article{Fray2009,
  title = "{Sublimation of ices of astrophysical interest: A bibliographic review}",
  volume = {57},
  ISSN = {0032-0633},
  DOI = {10.1016/j.pss.2009.09.011},
  number = {14–15},
  journal = {PSS},
  publisher = {Elsevier BV},
  author = {Fray,  N. and Schmitt,  B.},
  year = {2009},
  pages = {2053–2080}
}

@article {Hourdin1993,
      author = "Frédéric  Hourdin and Phu  Le Van and François  Forget and Olivier  Talagrand",
      title = "Meteorological Variability and the Annual Surface Pressure Cycle on {{M}ars}",
      journal = "JAS",
      year = "1993",
      publisher = "American Meteorological Society",
      address = "Boston MA, USA",
      volume = "50",
      number = "21",
      doi = "10.1175/1520-0469(1993)050<3625:MVATAS>2.0.CO;2",
      pages=      "3625 - 3640",
}

@article{Hourdin1995,
author = {Hourdin, Frédéric and Forget, François and Talagrand, Olivier},
title = {The sensitivity of the Martian surface pressure and atmospheric mass budget to various parameters: A comparison between numerical simulations and {Viking} observations},
journal = {J. Geophys. Res. (Planets)},
volume = {100},
number = {E3},
pages = {5501-5523},
doi = {10.1029/94JE03079},
year = {1995}
}

@article{Bertrand2017,
  title = "{3D modeling of organic haze in Pluto’s atmosphere}",
  volume = {287},
  ISSN = {0019-1035},
  DOI = {10.1016/j.icarus.2017.01.016},
  journal = {Icarus},
  publisher = {Elsevier BV},
  author = {Bertrand,  Tanguy and Forget,  Fran\c{c}ois},
  year = {2017},
  pages = {72–86}
}

@article{Spiga2008,
  title = "{Fast and accurate estimation of solar irradiance on Martian slopes}",
  volume = {35},
  ISSN = {1944-8007},
  DOI = {10.1029/2008gl034956},
  number = {15},
  journal = {GRL},
  publisher = {American Geophysical Union (AGU)},
  author = {Spiga,  Aymeric and Forget,  Fran\c{c}ois},
  year = {2008},
}

@article{Schenk2018,
  title = "{Basins,  fractures and volcanoes: Global cartography and topography of Pluto from New Horizons}",
  volume = {314},
  ISSN = {0019-1035},
  DOI = {10.1016/j.icarus.2018.06.008},
  journal = {Icarus},
  publisher = {Elsevier BV},
  author = {Schenk,  Paul Michael and Beyer,  Ross A. and McKinnon,  William B. and Moore,  Jeffrey M. and Spencer,  John R. and White,  Oliver L. and Singer,  Kelsi and Nimmo,  Francis and Thomason,  Carver and Lauer,  Tod R. and Robbins,  Stuart and Umurhan,  Orkan M. and Grundy,  William M. and Stern,  S. Alan and Weaver,  Harold A. and Young,  Leslie A. and Smith,  K. Ennico and Olkin,  Cathy},
  year = {2018},
  pages = {400–433}
}

@article{Schenk2021,
  title = "{Triton: Topography and Geology of a Probable Ocean World with Comparison to Pluto and Charon}",
  volume = {13},
  ISSN = {2072-4292},
  DOI = {10.3390/rs13173476},
  number = {17},
  journal = {Remote Sensing},
  publisher = {MDPI AG},
  author = {Schenk,  Paul and Beddingfield,  Chloe and Bertrand,  Tanguy and Bierson,  Carver and Beyer,  Ross and Bray,  Veronica and Cruikshank,  Dale and Grundy,  William and Hansen,  Candice and Hofgartner,  Jason and Martin,  Emily and McKinnon,  William and Moore,  Jeffrey and Robbins,  Stuart and Runyon,  Kirby and Singer,  Kelsi and Spencer,  John and Stern,  S. and Stryk,  Ted},
  year = {2021},
  pages = {3476}
}

@article{Hofgartner2023,
  title = "{Bolometric Hemispherical Albedo Map of Pluto from New Horizons Observations}",
  volume = {4},
  ISSN = {2632-3338},
  DOI = {10.3847/psj/ace3ab},
  number = {7},
  journal = {PSJ},
  publisher = {American Astronomical Society},
  author = {Hofgartner,  Jason D. and Buratti,  Bonnie J. and Beyer,  Ross A. and Ennico,  Kimberly and Grundy,  Will M. and Howett,  Carly J. A. and Johnson,  Perianne E. and Lauer,  Tod R. and Olkin,  Catherine B. and Spencer,  John R. and Alan Stern,  S. and Weaver,  Harold A. and Young,  Leslie A.},
  year = {2023},
  pages = {132}
}

@article{Weaver2008,
  title = "{Overview of the New Horizons Science Payload}",
  volume = {140},
  ISSN = {1572-9672},
  DOI = {10.1007/s11214-008-9376-6},
  number = {1–4},
  journal = {SSR},
  publisher = {Springer Science and Business Media LLC},
  author = {Weaver,  H. A. and Gibson,  W. C. and Tapley,  M. B. and Young,  L. A. and Stern,  S. A.},
  year = {2008},
  pages = {75–91}
}

@article{Gabasova2021,
  title = "{Global compositional cartography of Pluto from intensity-based registration of LEISA data}",
  volume = {356},
  ISSN = {0019-1035},
  DOI = {10.1016/j.icarus.2020.113833},
  journal = {Icarus},
  publisher = {Elsevier BV},
  author = {Gabasova,  L.R. and Schmitt,  B. and Grundy,  W. and Bertrand,  T. and Olkin,  C.B. and Spencer,  J.R. and Young,  L.A. and Ennico,  K. and Weaver,  H.A. and Stern,  S.A.},
  year = {2021},
  pages = {113833}
}

@article{Meza2019,
  title = "{Lower atmosphere and pressure evolution on Pluto from ground-based stellar occultations,  1988–2016}",
  volume = {625},
  ISSN = {1432-0746},
  DOI = {10.1051/0004-6361/201834281},
  journal = {A \& A},
  publisher = {EDP Sciences},
  author = {Meza,  E. and Sicardy,  B. and Assafin,  M. and Ortiz,  J. L. and Bertrand,  T. and Lellouch,  E. and Desmars,  J. and Forget,  F. and Bérard,  D. and Doressoundiram,  A. and Lecacheux,  J. and Oliveira,  J. Marques and Roques,  F. and Widemann,  T. and Colas,  F. and Vachier,  F. and Renner,  S. and Leiva,  R. and Braga-Ribas,  F. and Benedetti-Rossi,  G. and Camargo,  J. I. B. and Dias-Oliveira,  A. and Morgado,  B. and Gomes-Júnior,  A. R. and Vieira-Martins,  R. and Behrend,  R. and Tirado,  A. Castro and Duffard,  R. and Morales,  N. and Santos-Sanz,  P. and Jelínek,  M. and Cunniffe,  R. and Querel,  R. and Harnisch,  M. and Jansen,  R. and Pennell,  A. and Todd,  S. and Ivanov,  V. D. and Opitom,  C. and Gillon,  M. and Jehin,  E. and Manfroid,  J. and Pollock,  J. and Reichart,  D. E. and Haislip,  J. B. and Ivarsen,  K. M. and LaCluyze,  A. P. and Maury,  A. and Gil-Hutton,  R. and Dhillon,  V. and Littlefair,  S. and Marsh,  T. and Veillet,  C. and Bath,  K.-L. and Beisker,  W. and Bode,  H.-J. and Kretlow,  M. and Herald,  D. and Gault,  D. and Kerr,  S. and Pavlov,  H. and Faragó,  O. and Kl\"{o}s,  O. and Frappa,  E. and Lavayssière,  M. and Cole,  A. A. and Giles,  A. B. and Greenhill,  J. G. and Hill,  K. M. and Buie,  M. W. and Olkin,  C. B. and Young,  E. F. and Young,  L. A. and Wasserman,  L. H. and Devogèle,  M. and French,  R. G. and Bianco,  F. B. and Marchis,  F. and Brosch,  N. and Kaspi,  S. and Polishook,  D. and Manulis,  I. and Ait Moulay Larbi,  M. and Benkhaldoun,  Z. and Daassou,  A. and El Azhari,  Y. and Moulane,  Y. and Broughton,  J. and Milner,  J. and Dobosz,  T. and Bolt,  G. and Lade,  B. and Gilmore,  A. and Kilmartin,  P. and Allen,  W. H. and Graham,  P. B. and Loader,  B. and McKay,  G. and Talbot,  J. and Parker,  S. and Abe,  L. and Bendjoya,  Ph. and Rivet,  J.-P. and Vernet,  D. and Di Fabrizio,  L. and Lorenzi,  V. and Magazzú,  A. and Molinari,  E. and Gazeas,  K. and Tzouganatos,  L. and Carbognani,  A. and Bonnoli,  G. and Marchini,  A. and Leto,  G. and Sanchez,  R. Zanmar and Mancini,  L. and Kattentidt,  B. and Dohrmann,  M. and Guhl,  K. and Rothe,  W. and Walzel,  K. and Wortmann,  G. and Eberle,  A. and Hampf,  D. and Ohlert,  J. and Krannich,  G. and Murawsky,  G. and G\"{a}hrken,  B. and Gloistein,  D. and Alonso,  S. and Román,  A. and Communal,  J.-E. and Jabet,  F. and deVisscher,  S. and Sérot,  J. and Janik,  T. and Moravec,  Z. and Machado,  P. and Selva,  A. and Perelló,  C. and Rovira,  J. and Conti,  M. and Papini,  R. and Salvaggio,  F. and Noschese,  A. and Tsamis,  V. and Tigani,  K. and Barroy,  P. and Irzyk,  M. and Neel,  D. and Godard,  J. P. and Lanoiselée,  D. and Sogorb,  P. and Vérilhac,  D. and Bretton,  M. and Signoret,  F. and Ciabattari,  F. and Naves,  R. and Boutet,  M. and De Queiroz,  J. and Lindner,  P. and Lindner,  K. and Enskonatus,  P. and Dangl,  G. and Tordai,  T. and Eichler,  H. and Hattenbach,  J. and Peterson,  C. and Molnar,  L. A. and Howell,  R. R.},
  year = {2019},
  pages = {A42}
}

@article{Bertrand2022,
  title = "{Volatile transport modeling on Triton with new observational constraints}",
  volume = {373},
  ISSN = {0019-1035},
  DOI = {10.1016/j.icarus.2021.114764},
  journal = {Icarus},
  publisher = {Elsevier BV},
  author = {Bertrand,  T. and Lellouch,  E. and Holler,  B.J. and Young,  L.A. and Schmitt,  B. and Marques Oliveira,  J. and Sicardy,  B. and Forget,  F. and Grundy,  W.M. and Merlin,  F. and Vangvichith,  M. and Millour,  E. and Schenk,  P.M. and Hansen,  C.J. and White,  O.L. and Moore,  J.M. and Stansberry,  J.A. and Oza,  A.V. and Dubois,  D. and Quirico,  E. and Cruikshank,  D.P.},
  year = {2022},
  pages = {114764}
}

@article{Lellouch2015,
  title = "{Exploring the spatial,  temporal,  and vertical distribution of methane in Pluto’s atmosphere}",
  volume = {246},
  ISSN = {0019-1035},
  DOI = {10.1016/j.icarus.2014.03.027},
  journal = {Icarus},
  publisher = {Elsevier BV},
  author = {Lellouch,  E. and de Bergh,  C. and Sicardy,  B. and Forget,  F. and Vangvichith,  M. and K\"{a}ufl,  H.-U.},
  year = {2015},
  pages = {268–278}
}

@article{Gladstone2016,
  title = "{The atmosphere of Pluto as observed by New Horizons}",
  volume = {351},
  ISSN = {1095-9203},
  DOI = {10.1126/science.aad8866},
  number = {6279},
  journal = {Science},
  publisher = {American Association for the Advancement of Science (AAAS)},
  author = {Gladstone,  G. Randall and Stern,  S. Alan and Ennico,  Kimberly and Olkin,  Catherine B. and Weaver,  Harold A. and Young,  Leslie A. and Summers,  Michael E. and Strobel,  Darrell F. and Hinson,  David P. and Kammer,  Joshua A. and Parker,  Alex H. and Steffl,  Andrew J. and Linscott,  Ivan R. and Parker,  Joel Wm. and Cheng,  Andrew F. and Slater,  David C. and Versteeg,  Maarten H. and Greathouse,  Thomas K. and Retherford,  Kurt D. and Throop,  Henry and Cunningham,  Nathaniel J. and Woods,  William W. and Singer,  Kelsi N. and Tsang,  Constantine C. C. and Schindhelm,  Rebecca and Lisse,  Carey M. and Wong,  Michael L. and Yung,  Yuk L. and Zhu,  Xun and Curdt,  Werner and Lavvas,  Panayotis and Young,  Eliot F. and Tyler,  G. Leonard and Bagenal,  F. and Grundy,  W. M. and McKinnon,  W. B. and Moore,  J. M. and Spencer,  J. R. and Andert,  T. and Andrews,  J. and Banks,  M. and Bauer,  B. and Bauman,  J. and Barnouin,  O. S. and Bedini,  P. and Beisser,  K. and Beyer,  R. A. and Bhaskaran,  S. and Binzel,  R. P. and Birath,  E. and Bird,  M. and Bogan,  D. J. and Bowman,  A. and Bray,  V. J. and Brozovic,  M. and Bryan,  C. and Buckley,  M. R. and Buie,  M. W. and Buratti,  B. J. and Bushman,  S. S. and Calloway,  A. and Carcich,  B. and Conard,  S. and Conrad,  C. A. and Cook,  J. C. and Cruikshank,  D. P. and Custodio,  O. S. and Ore,  C. M. Dalle and Deboy,  C. and Dischner,  Z. J. B. and Dumont,  P. and Earle,  A. M. and Elliott,  H. A. and Ercol,  J. and Ernst,  C. M. and Finley,  T. and Flanigan,  S. H. and Fountain,  G. and Freeze,  M. J. and Green,  J. L. and Guo,  Y. and Hahn,  M. and Hamilton,  D. P. and Hamilton,  S. A. and Hanley,  J. and Harch,  A. and Hart,  H. M. and Hersman,  C. B. and Hill,  A. and Hill,  M. E. and Holdridge,  M. E. and Horanyi,  M. and Howard,  A. D. and Howett,  C. J. A. and Jackman,  C. and Jacobson,  R. A. and Jennings,  D. E. and Kang,  H. K. and Kaufmann,  D. E. and Kollmann,  P. and Krimigis,  S. M. and Kusnierkiewicz,  D. and Lauer,  T. R. and Lee,  J. E. and Lindstrom,  K. L. and Lunsford,  A. W. and Mallder,  V. A. and Martin,  N. and McComas,  D. J. and McNutt,  R. L. and Mehoke,  D. and Mehoke,  T. and Melin,  E. D. and Mutchler,  M. and Nelson,  D. and Nimmo,  F. and Nunez,  J. I. and Ocampo,  A. and Owen,  W. M. and Paetzold,  M. and Page,  B. and Pelletier,  F. and Peterson,  J. and Pinkine,  N. and Piquette,  M. and Porter,  S. B. and Protopapa,  S. and Redfern,  J. and Reitsema,  H. J. and Reuter,  D. C. and Roberts,  J. H. and Robbins,  S. J. and Rogers,  G. and Rose,  D. and Runyon,  K. and Ryschkewitsch,  M. G. and Schenk,  P. and Sepan,  B. and Showalter,  M. R. and Soluri,  M. and Stanbridge,  D. and Stryk,  T. and Szalay,  J. R. and Tapley,  M. and Taylor,  A. and Taylor,  H. and Umurhan,  O. M. and Verbiscer,  A. J. and Versteeg,  M. H. and Vincent,  M. and Webbert,  R. and Weidner,  S. and Weigle,  G. E. and White,  O. L. and Whittenburg,  K. and Williams,  B. G. and Williams,  K. and Williams,  S. and Zangari,  A. M. and Zirnstein,  E.},
  year = {2016},
}

@article{Hinson2017,
  title = {Radio occultation measurements of Pluto’s neutral atmosphere with New Horizons},
  volume = {290},
  ISSN = {0019-1035},
  DOI = {10.1016/j.icarus.2017.02.031},
  journal = {Icarus},
  publisher = {Elsevier BV},
  author = {Hinson,  D.P. and Linscott,  I.R. and Young,  L.A. and Tyler,  G.L. and Stern,  S.A. and Beyer,  R.A. and Bird,  M.K. and Ennico,  K. and Gladstone,  G.R. and Olkin,  C.B. and P\"{a}tzold,  M. and Schenk,  P.M. and Strobel,  D.F. and Summers,  M.E. and Weaver,  H.A. and Woods,  W.W.},
  year = {2017},
  pages = {96–111}
}

@misc{Lange2025plutodata,
  doi = {10.7910/DVN/3BBLG6},
  author = {Lange,  Lucas and Bertrand,  Tanguy and Belissa,  Victor and Capry,  Sabrina and Young,  Leslie A and Falco,  Aurelien},
  title = "{Replication Data for: Replication Data for: Modeling the formation of N$_2$ and CH$_4$ frost on Pluto's slopes [Dataset]}",
  publisher = {Harvard Dataverse},
  year = {2025}
}

@article{Earle2017,
  title = {Long-term surface temperature modeling of Pluto},
  volume = {287},
  ISSN = {0019-1035},
  DOI = {10.1016/j.icarus.2016.09.036},
  journal = {Icarus},
  publisher = {Elsevier BV},
  author = {Earle,  Alissa M. and Binzel,  Richard P. and Young,  Leslie A. and Stern,  S.A. and Ennico,  K. and Grundy,  W. and Olkin,  C.B. and Weaver,  H.A.},
  year = {2017},
  pages = {37–46}
}

@article{Mishra2025,
  title = "{Investigating the Extent of Bladed Terrain on Pluto via Photometric Surface Roughness}",
  volume = {130},
  ISSN = {2169-9100},
  url = {10.1029/2024JE008554},
  DOI = {10.1029/2024je008554},
  number = {7},
  journal = {J. Geophys. Res. (Planets)},
  publisher = {American Geophysical Union (AGU)},
  author = {Mishra,  I. and Dhingra,  R. and Buratti,  B. J. and Seignovert,  B. and White,  O. L.},
  year = {2025},
}

@article{Protopapa2020,
  title = "{Disk-resolved Photometric Properties of Pluto and the Coloring Materials across its Surface}",
  volume = {159},
  ISSN = {1538-3881},
  url = {10.3847/1538-3881/ab5e82},
  DOI = {10.3847/1538-3881/ab5e82},
  number = {2},
  journal = {AJ},
  publisher = {American Astronomical Society},
  author = {Protopapa,  Silvia and Olkin,  Cathy B. and Grundy,  Will M. and Li,  Jian-Yang and Verbiscer,  Anne and Cruikshank,  Dale P. and Gautier,  Thomas and Quirico,  Eric and Cook,  Jason C. and Reuter,  Dennis and Howett,  Carly J. A. and Stern,  Alan and Beyer,  Ross A. and Porter,  Simon and Young,  Leslie A. and Weaver,  Hal A. and Ennico,  Kim and Dalle Ore,  Cristina M. and Scipioni,  Francesca and Singer,  Kelsi},
  year = {2020},
  pages = {74}
}

@INCOLLECTION{Dobrovolskis1997,
       author = {{Dobrovolskis}, A.~R. and {Peale}, S.~J. and {Harris}, A.~W.},
        title = "{Dynamics of the Pluto-Charon Binary}",
    booktitle = {Pluto and Charon},
         year = 1997,
       editor = {{Stern}, S. Alan and {Tholen}, David J.},
        pages = {159},
}
\onecolumn

\begin{appendix}
\section{Comparison between New-Horizons observations and model outputs using a CH$_4$ ice albedo of 0.67}

    \begin{figure*}[h!] \centering \includegraphics[width=0.9\textwidth]{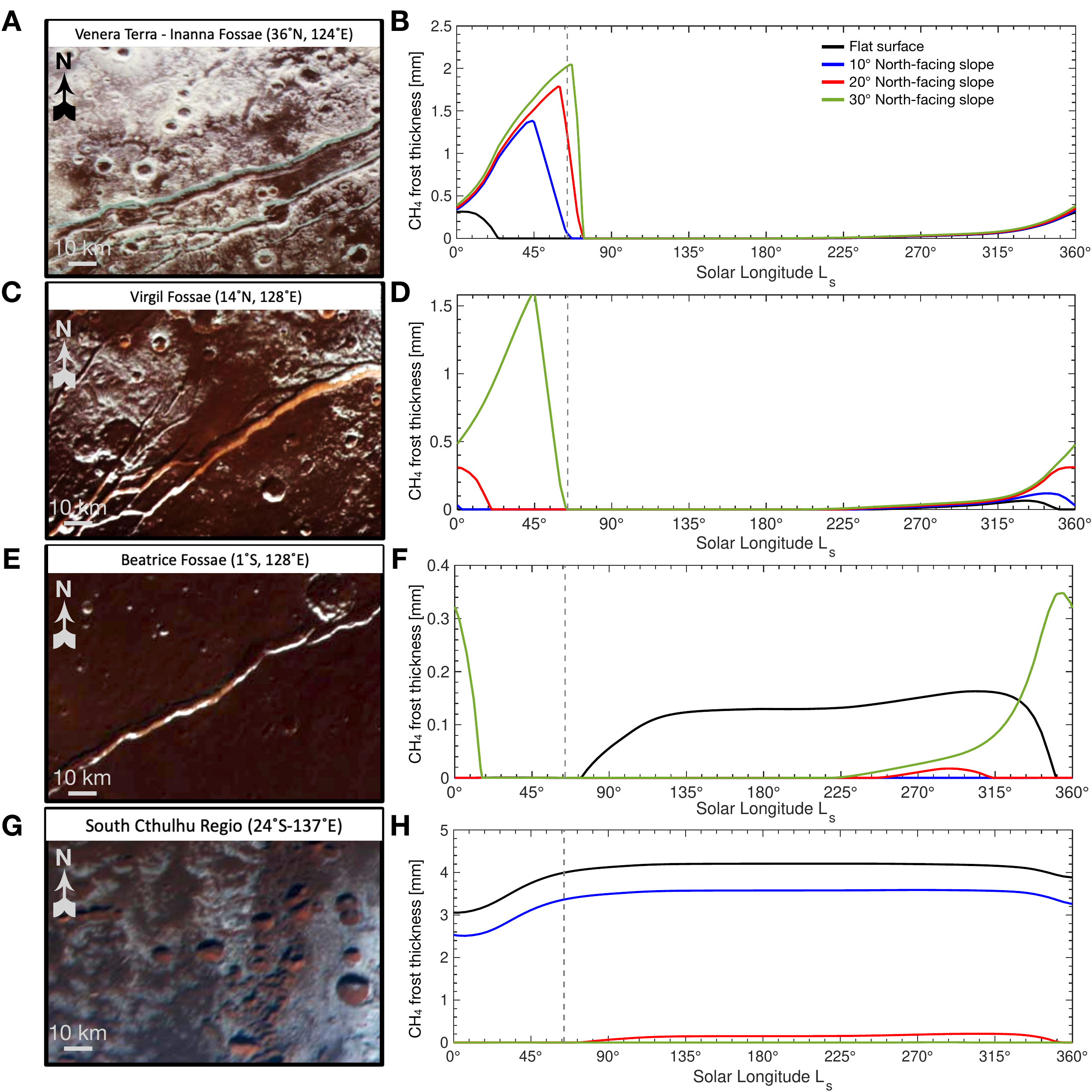} \caption{Comparisons between frost observations on Ralph/MVIC color images in the Cthulhu region reported by \cite{Bertrand2020} at Venera Terra (A), Virgil Fossae (C), Beatrice Fossae (E), and the southern part of the region (G) and the model outputs (right) assuming a baseline albedo  of 0.67 for methane ice. This figure is the analog of Fig. \ref{fig:compbertrand_069}, with all model parameters kept identical except that the albedo of CH$_4$ ice was set to its baseline value of 0.67 instead of 0.69. } 
    \label{fig:compbertrand_067}
    \end{figure*}
\end{appendix}

\end{document}